\documentclass[pdftex]{mn2e}
\usepackage{epsf,graphicx}

\newif\ifAMStwofonts
\newcommand{\Mach} {M}

\newcommand{\vp} {v_{\phi 0}}
\newcommand{\vr} {v_{R 0}}
\newcommand{\dP} {\Phi_1}

\newcommand{\xim} {\left(\frac{1-\xi}{2}\right)}
\newcommand{\xip} {\left(\frac{1+\xi}{2}\right)}

\newcommand{\sumj} {\sum_{j=|k|}^{\infty}}

\def\spose#1{\hbox to 0pt{#1\hss}}
\def\lta{\mathrel{\spose{\lower 3.2pt\hbox{$\sim$}}
    \raise 1.2pt\hbox{$<$}}}
\def\gta{\mathrel{\spose{\lower 3.2pt\hbox{$\sim$}}
    \raise 1.2pt\hbox{$>$}}}

\title{Unstable modes of non-axisymmetric gaseous discs}
\author[Naser M. Asghari and Mir Abbas Jalali]
       {Naser M. Asghari$^{1,2}$\thanks{naser\_ma@iasbs.ac.ir}
        and Mir Abbas Jalali$^3$\thanks{mjalali@sharif.edu} \\
        $^1$Institute for Advanced Studies in Basic Sciences,
        P.O. Box 45195-1159, Zanjan, IRAN \\
        $^2$Aerospace Research Institute,
        P.O. Box 15875-3885, Tehran, IRAN \\
        $^3$Department of Mechanical Engineering,
            Sharif University of Technology, P.O. Box 11365-9567, Tehran, IRAN}
\date{}

\pagerange{\pageref{firstpage}--\pageref{lastpage}}
\pubyear{2006}

\begin{document}

\maketitle

\label{firstpage}

\begin{abstract}
We present a perturbation theory for studying the instabilities of
non-axisymmetric gaseous discs. We perturb the dynamical equations
of self-gravitating fluids in the vicinity of a non-axisymmetric
equilibrium, and expand the perturbed physical quantities in terms
of a complete basis set and a small non-axisymmetry parameter
$\epsilon$. We then derive a linear eigenvalue problem in matrix
form, and determine the pattern speed, growth rate and mode shapes
of the first three unstable modes. In non-axisymmetric
discs the amplitude and phase angle of travelling waves are
functions of both the radius $R$ and the azimuthal angle $\phi$.
This is due to the interaction of different wave components in the
response spectrum. We demonstrate that wave interaction in unstable
discs with small initial asymmetries, can develop dense clumps
during the phase of exponential growth. Local clumps, which occur on
the major spiral arms, can constitute seeds of gas giant planets in
accretion discs.
\end{abstract}

\begin{keywords} accretion discs -- hydrodynamics -- instabilities --
galaxies: structure -- galaxies: spiral
\end{keywords}

\section{Introduction}
Unstable density waves inside self-gravitating gaseous media play an
important role in structuring of spiral galaxies and accretion
discs. Prior works in the literature deal with the instability of
differentially rotating discs, which are initially circular. Aoki et
al. (1979) carried out a global analysis of spiral modes in the
soft-centred, gaseous Kuzmin (1956) disc using Hunter's (1963)
device. Lemos et al. (1991) studied the unstable axisymmetric modes
of the more puzzling scale-free discs and solved the dispersion
relation to determine the sufficient conditions for ring formation
and for the existence of breathing modes. Recently Goodman \& Evans
(1999, hereafter GE) constructed the normal modes of the scale-free
Mestel (1963) disc in the Cowling approximation. For
self-gravitating modes they computed the ratio of growth rate to
pattern speed by solving a recurrence relation in the
Mellin-transform space. Density waves in the presence of an external
potential field, like that of a massive central object, have also
been extensively studied using semi-analytical and finite difference
techniques (e.g., Papaloizou \& Lin 1989; Savonije \& Heemskerk
1990; Papaloizou \& Savonije 1991).

The assumption of axisymmetry of the equilibrium (or initial) state,
however, imposes significant simplifications on the solution of
dynamical equations. In fact, unstable modes of circular discs are
decoupled because the only $\phi$-dependent term, $\exp({\rm
i}m\phi)$ (${\rm i}=\sqrt{-1}$), is factored out of the linearized
Poisson and hydrodynamic equations. Here $\phi$ is the azimuthal
angle and $m$ is the wave number. Such a factorization is impossible
if the unperturbed state is non-axisymmetric. Azimuthal modes of
perturbed non-axisymmetric discs can interact with the Fourier
components of the equilibrium state and result in more complex
patterns.

In this paper we attempt to formulate the instability problem of
non-axisymmetric discs based on a perturbation theory, and present a
mechanism for the generation of dense clumps through {\it wave
interaction} in {\it linear regime}. That occurs when a
non-axisymmetric disc is destabilized. We apply our formulation to
the simple discs of Jalali \& Abolghasemi (2002, hereafter JA),
which reduce to Mestel's disc in axisymmetric limit. We have chosen
JA models as our unperturbed state because of two reasons. Firstly,
closed form (exact) expressions are available for physical
quantities like the potential--density pair, pressure, sound speed
and velocity components. Secondly, both the potential and the
surface density have a single Fourier component. This minimizes the
number of interacting terms in our subsequent analysis.

We review steady-state, non-axisymmetric, scale-free discs of JA in
section \ref{sec:non-axi-symmetric-discs} and introduce a cutout
function that helps us handle the singularity at the centre. We
present a first-order perturbation theory for non-axisymmetric
gaseous discs in section \ref{sec:perturbation-formulation}, where
we also introduce a basis set by which we expand the perturbed
physical quantities. We then derive an eigenvalue problem for
determining the normal modes of non-axisymmetric discs and discuss
on the algorithms used for solving the resulting eigensystem.
In sections \ref{sec::modes-of-circular-discs} and
\ref{sec:modes-nonaxi-discs}, we study the unstable modes of
circular and non-axisymmetric discs for the fundamental wave
numbers $m=0$, 1 and 2, and discuss on the results and their
physical implications in section \ref{sec:discussion}.

\begin{figure}
\centerline{\hbox{\epsfxsize=8.5truecm\epsfbox{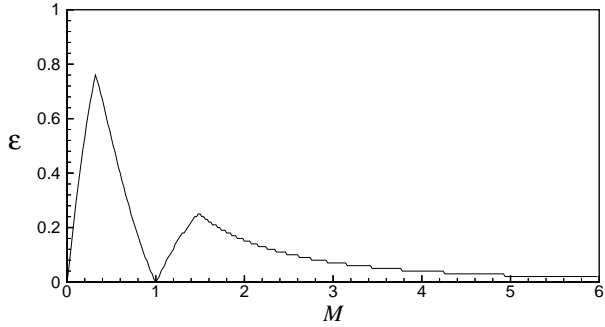}}}
\caption{The region under the solid line is the allowable zone in
the parameter space of a bisymmetric disc with $n=2$.}
\label{fig:fig1}
\end{figure}

\section{Non-axisymmetric gaseous discs in equilibrium}
\label{sec:non-axi-symmetric-discs}
There are few papers in the
literature that address non-axisymmetric equilibria of gaseous discs
(Syer \& Tremaine 1996; Galli et al. 2001; JA). In this section we
briefly review one of {\it exact} non-axisymmetric models of JA
whose surface density consists of a simple harmonic function of the
azimuthal angle.

\subsection{Basic derivations}
\label{sec::JA-models}

We define $(R,\phi)$ as the usual polar coordinates and assume a
self-gravitating, inviscid, compressible fluid in a steady two
dimensional state so that the surface density distribution and its
self-consistent potential are scale-free as
\begin{eqnarray}
\Sigma \left(R,\phi \right )&=& \Sigma_s \frac{R_0}{R}f(\phi),
\label{eq:equilibrium-density} \\
\Phi \left( R,\phi \right ) &=& v_0^2 \left[ \ln
\frac{R}{R_0}+\frac{1}{2}g\left(\phi\right)\right],~~v_0^2=2\pi
G\Sigma_s R_0. \label{eq:equilibrium-potential}
\end{eqnarray}
Here $G$ is the constant of gravitation, and $R_0$ and $\Sigma_s$
are the normalizing constants of the length and surface density,
respectively. Our discs are non-axisymmetric because $\Sigma$ and
$\Phi$ depend on the azimuthal angle $\phi$. Due to the continuity
of physical quantities in the azimuthal direction, the functions
$f(\phi)$ and $g(\phi)$ are $2\pi$-periodic in $\phi$. For the sake
of simplicity, we work with non-axisymmetric discs of the form
\begin{equation}
f\left(\phi\right) = 1+\epsilon \cos n\phi, \label{eq:simple-disc}
\end{equation}
where $0\leq \epsilon <1$. The $\phi$-dependent part of the
potential reads
\begin{equation}
g\left(\phi\right) =- \frac{2}{n}\epsilon \cos n\phi.
\label{eq:simple potential}
\end{equation}
For $\epsilon = 0$, our discs reduce to the isothermal, gaseous
Mestel disc. Let us respectively define $c_0$, $V_c$ and
$\Mach$=$V_c/c_0$ as the sound speed, velocity of circular orbits
and Mach number in Mestel's disc. The components of the velocity
field associated with (\ref{eq:equilibrium-density}) and
(\ref{eq:equilibrium-potential}) are obtained after solving the
hydrodynamic equations as (JA)
\begin{eqnarray}
v_{\phi}(\phi) \!\!\!&=&\!\!\! \frac{V_{c}}{1+\epsilon \cos n\phi},
                \label{eq:azim-velocity-for-n2} \\
v_{R}(\phi) \!\!\!&=&\!\!\! -\epsilon
\frac{1+\Mach^2}{2\Mach^2}V_{c} \left(\frac{2\sin n\phi}{1+n}
   -\frac{\epsilon \sin 2n\phi}{4n^2-1} \right ).
\label{eq:rad-velocity-for-n2}
\end{eqnarray}
The sound speed $c(\phi)$ and the pressure distribution $P(\phi)$
are then given by
\begin{eqnarray}
c(\phi) \!\!\!\!&=&\!\!\!\! c_0
\sqrt {{(1+\Mach^2)A(\phi)-2\Mach ^2 \over
        2\left ( 1+\epsilon \cos n\phi \right )^2}},
\label{eq:sound-speed-for-n2} \\
P(\phi) \!\!\!\! &=& \!\!\!\! c^2 \Sigma, \label{eq:nonaxi-pressure} \\
A(\phi) \!\!\!\!&=&\!\!\!\! 2\left(1+\epsilon\cos n\phi\right) \left
(1+ \frac{\epsilon \cos n\phi}{1+n} +\frac{n\epsilon^2 \cos
2n\phi}{4n^2-1} \right ). \label{eq:function-A-in-c}
\end{eqnarray}
$c$ reduces to $c_0$ in the limit of $\epsilon=0$. From
(\ref{eq:equilibrium-potential}) and (\ref{eq:nonaxi-pressure}), and
equation (2) in JA, one can verify that $v_0^2$=$V_c^2+c_0^2$. Sound
waves are determinate if the pressure is positive, i.e., $c^2
>0$. A necessary and sufficient condition for the positiveness of
$P$ is
\begin{equation}
0 < \frac{2\Mach^2}{1+\Mach^2}<A_{\rm min}.
\label{con:positivespeed}
\end{equation}
Here $A_{\rm min}$ is the minimum of $ A(\phi)$ over the interval
$0\le \phi \le 2\pi$. Inequality (\ref{con:positivespeed}) restricts
model parameters. JA models allow for transonic flows because the
sound speed depends on $\phi$ and it varies along a given streamline.
In principle, transition from subsonic to supersonic speeds can take
place smoothly but the reverse phenomenon is impossible without a
shock wave. To avoid shock waves, we follow JA and exclude transonic
flows from our calculations and investigate supersonic and subsonic
discs.

In this paper we choose a bisymmetric equilibrium with $n=2$. The
allowable region of the parameter space $(\Mach,\epsilon)$ is
determined based on two requirements. Firstly the square of the
sound speed must be positive. Secondly, transonic flows are
excluded. Fig. \ref{fig:fig1} shows the admissible zone of the
parameter space. Note that for $\epsilon >0$ the Mach number is a
function of the azimuthal angle:
\begin{equation}
\textbf{M}(M,\epsilon,\phi)={\sqrt{v_R^2+v_\phi^2}\over c(\phi)}.
\end{equation}
In the limit of circular discs ($\epsilon=0$) we obtain
$\textbf{M}=M$ as expected.

\subsection{Active surface density}
\label{sec:cutout-definition} We are studying scale-free gaseous
discs whose surface density and potential diverge near the center
according to (\ref{eq:equilibrium-density}) and
(\ref{eq:equilibrium-potential}). The linear stability problem of
scale-free discs is ill-posed unless one decides on the way that
inward traveling waves are reflected off the cusp. Several methods
have been devised to tackle this problem. For scale-free stellar
discs Zang (1976) (see also Evans \& Read 1998a,b) used an inner
cutout, which carves out the density function and deactivates the
central region of the stellar system. In this way, stars with high
orbital frequencies do not participate in perturbations. For the
gaseous Mestel disc GE used a hard internal boundary that fixes the
phase with which waves are reflected from the centre. In this paper
we apply Evans \& Read's (1998a) method to our cuspy gaseous discs
of \S\ref{sec::JA-models} and deactivate some parts of the disc mass
by the cutout function
\begin{equation}
H(R)={R^{N_{\rm in}}\over \left [ R^{N_{\rm in}}+(\alpha
R_0)^{N_{\rm in }}\right ]}{R_0^{N_{\rm out}}\over \left [(\gamma
R)^{N_{\rm out}}+R_0^{N_{\rm out}}\right ]}. \label{eq:cutout-fnc}
\end{equation}
We then analyze the perturbations of the {\it active} density
\begin{equation} \Sigma _{\rm act}(R)=\Sigma(R)H(R).
\label{eq:cutout-density}
\end{equation}
Here $R_{\rm in}=\alpha R_0$ and $R_{\rm out}=R_0/\gamma$ are the
inner and outer cutout radii, respectively. Since the cutout is not
applied to the potential, the immobile mass will resemble a rigid
bulge/halo component that does not participate in the disc dynamics,
but it contributes to the force field. Instability properties are
not sensitive to the choice of $R_{\rm out}$. We apply the outer
cutout for computational convenience and use a large $R_{\rm out}$
far beyond the region that instabilities occur. The nature of
functions by which we will expand perturbed fields, constrains the
feasible values of $N_{\rm in}$ and $N_{\rm out}$. We have used
$N_{\rm in}$= $N_{\rm out}=2$ throughout this paper.

\section{Perturbation theory of non-axisymmetric discs}
\label{sec:perturbation-formulation} We define $\textbf{u}=\left
(u_1,u_2,u_3 \right )^T\equiv (\Sigma,v_R,v_{\phi})^{T}$ and
consider the first order Eulerian perturbations of physical
quantities as
\begin{eqnarray}
\textbf{u} &\rightarrow&
\textbf{u}_0(R,\phi)+\textbf{u}_1(R,\phi,t),
\nonumber \\
\Phi &\rightarrow& \Phi_0(R,\phi)+\Phi _1(R,\phi,t),
\label{eq::perturbations-for-all} \\
P &\rightarrow&
c^2(\phi)\Sigma_0(R,\phi)+c^2(\phi)\Sigma_1(R,\phi,t). \nonumber
\end{eqnarray}
The superscript $T$ denotes transpose and the subscripts 0 and 1
refer to the equilibrium and perturbed states, respectively. We
assume that the perturbed quantities,
$\delta_1$$\equiv$$\left(\Phi_1,\Sigma_1,v_{R1},v_{\phi 1}
\right)^T$, are composed of a discrete set of normal modes, each of
the form
\begin{eqnarray}
\delta_{1,m}(R,\phi,t,\epsilon) \!\! &=& \!\! e^{-{\rm i} \omega_m
t}\delta^{(m)}_{1}(R,\phi,\epsilon), \label{eq::discrete-modes} \\
\omega_m \!\! &=& \!\! \Omega_m + {\rm i}s_m,~~{\rm i}=\sqrt{-1},
\label{eq::define-omega}
\end{eqnarray}
where $\omega_m$, $\Omega_m$ and $s_m$ are respectively the
eigenfrequency, pattern speed and growth rate of the $m$th mode. It
is remarked that the real part of $\delta_{1,m}$ gives the physical
solution.

We use the equilibrium fields of \S\ref{sec:non-axi-symmetric-discs}
and substitute from (\ref{eq::perturbations-for-all}) into the continuity
and momentum equations of self-gravitating fluids
[equations (1E-3) and (6-20) in Binney \& Tremaine 1987].
After collecting the first-order terms in perturbed variables,
we obtain
\begin{equation}
{\bf L}\cdot \textbf{u}_1 = -\left (0,\partial _R\dP,R^{-1}\partial
_\phi \dP \right )^T, \label{eq:linearized-perturbed-equations}
\end{equation}
with $\partial _\mu=\partial /\partial \mu$. The elements of the
linear operator ${\bf L}=[L_{ij}]$, ($i,j=1,2,3$), have been given
in Appendix A. $\Phi_1$ and $\Sigma_1$ should also satisfy Poisson's
integral
\begin{equation}
\Phi_1=-G\int \int {\Sigma_1 R'{\rm d}R'{\rm d}\phi'
\over \sqrt{R^2+R'^2-2RR'\cos(\phi-\phi')}}.
\label{eq::Poisson-integral-equation}
\end{equation}

Modes of non-axisymmetric discs bifurcate from those of circular
discs once we let $\epsilon$ grow from zero. Therefore, we seek for
a perturbation solution for $\delta_{1,m}$ in terms of the
non-axisymmetry parameter $\epsilon$. Moreover, continuity of
physical quantities in the azimuthal direction implies the
periodicity conditions
\begin{eqnarray}
\textbf{u}_0(R,\phi+2\pi) \!\!\! &=& \!\!\! \textbf{u}_0(R,\phi), \\
\delta^{(m)}_1(R,\phi+2\pi,\epsilon) \!\!\! &=& \!\!\!
\delta^{(m)}_1(R,\phi,\epsilon).
\end{eqnarray}
These requirements suggest to represent $\textbf{u}_0$ by the
Fourier series
\begin{equation}
\textbf{u}_0(R,\phi) = \sum_{p=-\infty}^{+\infty}\epsilon ^{\vert
p\vert}e^{{\rm i}np\phi}\textbf{u}_{0,p}(R),
\label{eq::equilibrium-fields-expansion}
\end{equation}
and propose the {\it Ansatz}
\begin{equation}
\delta ^{(m)}_1(R,\phi,\epsilon) = \sum_{l=-\infty}^{+\infty}
\epsilon ^{\vert l\vert} e^{{\rm i}(m+nl)\phi} \hat\delta
^{(m+nl)}_1(R). \label{eq::perturbed-fields-expansion}
\end{equation}
Using equations (\ref{eq::equilibrium-fields-expansion}) and
(\ref{eq::perturbed-fields-expansion}) and substituting from
(\ref{eq::discrete-modes}) in
(\ref{eq:linearized-perturbed-equations}), we obtain the determining
equation of the $m$th mode as
\begin{eqnarray}
&-& \!\!\!\!\! \left ( {{\rm i}\omega_m R_0\over c_0 } \right )
\hat\textbf{u}^{(m+nl)}_1 \!\!+\!\! \sum_{p=-\infty}^{+\infty}
\epsilon^{\vert p\vert} {\bf D}_p \cdot
\hat\textbf{u}^{(m+nl-np)}_1 \nonumber \\
&=& -{R_0\over c_0}\left ( 0,\partial _R \hat\Phi^{(m+nl)}_1,{\rm i}
(m+nl)R^{-1}
\hat\Phi^{(m+nl)}_1 \right )^T, \label{eq::perturbed-vec-m-mode} \\
&{}& \qquad \qquad \qquad l=0,\pm 1,\pm 2,\ldots, \nonumber
\end{eqnarray}
which is a system of linear ordinary differential equations with
respect to $R$. Here ${\bf D}_p$s are linear differential operators
(they are square matrices of dimension $3\times 3$) whose elements
have been given in Appendix B for $0\le \vert p\vert \le 3$. The
second term on the left side of (\ref{eq::perturbed-vec-m-mode})
shows that how different azimuthal waves interact. The interacting
terms rapidly decay proportional to $\epsilon^{|p|}$ for $\epsilon
\ll 1$. In the axisymmetric limit $(\epsilon=0)$, equation
(\ref{eq::discrete-modes}) reduces to
\begin{equation}
\delta_{1,m}(R,\phi,t) = e^{-{\rm i}\omega_m t} e^{{\rm i}m\phi}
\hat\delta^{(m)}_1(R),
\end{equation}
so that $\omega_m$ is associated with a single azimuthal wave
component. Modal decomposition of this kind is impossible for
initially non-axisymmetric discs, for the equilibrium fields include
the powers of $e^{\pm {\rm i} n \phi}$ that cause wave interaction.

\subsection{Choice of basis functions}
\label{sec:basis-functions} One way for solving
(\ref{eq::Poisson-integral-equation}) and
(\ref{eq::perturbed-vec-m-mode}) is through expanding
$\hat\delta^{(k)}_1(R)$ [see equation
(\ref{eq::perturbed-fields-expansion})] in terms of some basis,
preferably biorthogonal, functions. Clutton-Brock (1972) found such
a set, which was then recalculated by Aoki \& Iye (1978) in terms of
associated Legendre functions. We follow them and expand the
functions $\hat\delta ^{(k)}_1(R)$ as
\begin{eqnarray}
\hat\Phi^{(k)}_1(R) \!\!\! &=& \!\!\! -2\pi G \Sigma_s R_C
\xim^{\frac 12} \sumj {\hat a^{k}_j \over 2j+1 }
\hat{P}^{|k|}_j(\xi), \label{eq:bi-orthogonal-potential}
\\
\hat\textbf{u}^{(k)}_1(R) \!\!\! &=& \!\!\! \sumj
\hat{P}^{|k|}_j(\xi) \textbf{W}\cdot \hat\textbf{X},
\label{eq:bi-orthogonal-u-vector}
\end{eqnarray}
where $\hat\textbf{X}$=$(\hat a^k_j,\hat b^k_j,\hat c^k_j)^T$ is a
to-be-determined vector of complex coefficients and
\begin{eqnarray}
\textbf{W} \!\!\! &=& \!\!\!
{\rm diag}\left [\Sigma_s w_1(\xi),{\rm i}c_0w_2(\xi),c_0 w_2(\xi)\right ], \\
\xi \!\!\! &=& \!\!\! {R^2-R_C^2\over R^2+R_C^2},~w_1=\left ( 1-\xi
\over 2\right )^{\frac 32},~w_2=\left ( 1-\xi \over 1+\xi \right
)^{\frac 12}. \nonumber
\end{eqnarray}
$\hat{P}^{|k|}_j \left(\xi\right)$ are the normalized associated
Legendre functions defined by
\begin{equation}
\hat{P}^{|k|}_j\left(\xi\right) = \left[ \frac{(j-|k|)!\
(2j+1)}{2(j+|k|)!}\right]^{\frac 12} P^{|k|}_j\left(\xi\right).
\end{equation}
We set the scale length of Clutton-Brock functions, $R_C$, to the
normalizing length of (\ref{eq:equilibrium-density}) and
(\ref{eq:equilibrium-potential}). i.e., $R_C=R_0$. The criterion in
choosing the {\it weight} function $w_2(\xi)$ was the convergence of
proceeding definite integrals that constitute the coefficients of
the matrix $\textbf{A}$ in (\ref{eq:eigenvalue-equation}). If $k
\neq 0$, the functions $\hat{P}^{|k|}_j(\xi )$ give the factor
$[(1+\xi)/2]^{|k|/2}$. Therefore, the expansions used for velocity
components are not singular at the origin (when $\xi \rightarrow
-1$). For $k=0$, the perturbed velocity components become singular
for $\xi=-1$. Nonetheless, the energy and angular momentum are not
singular there and the continuity equation is satisfied.

\begin{figure*}
\centerline{\hbox{\epsfxsize=5.5truecm\epsfbox{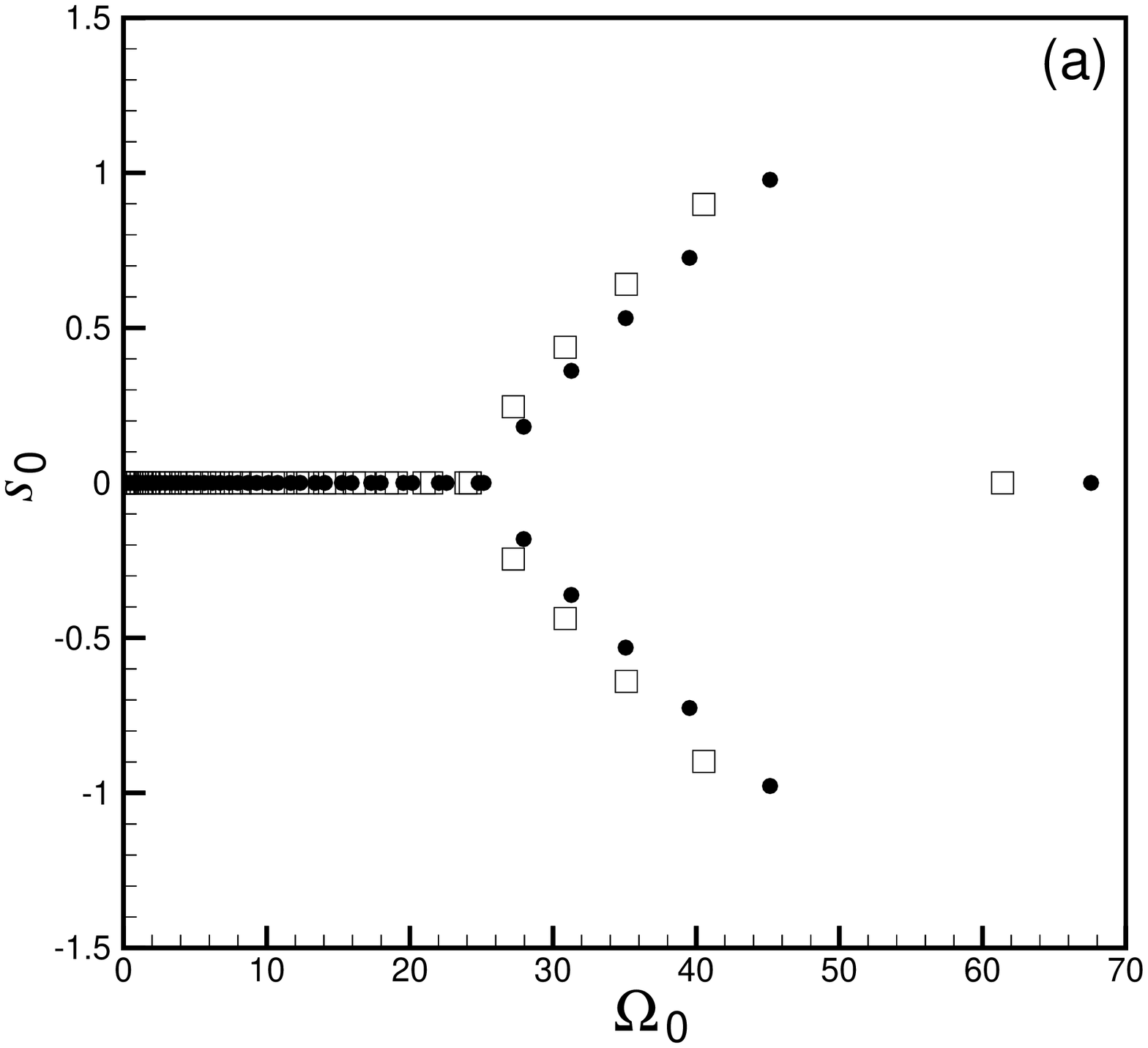}}
            \hspace{0.2truecm}
            \hbox{\epsfxsize=5.5truecm\epsfbox{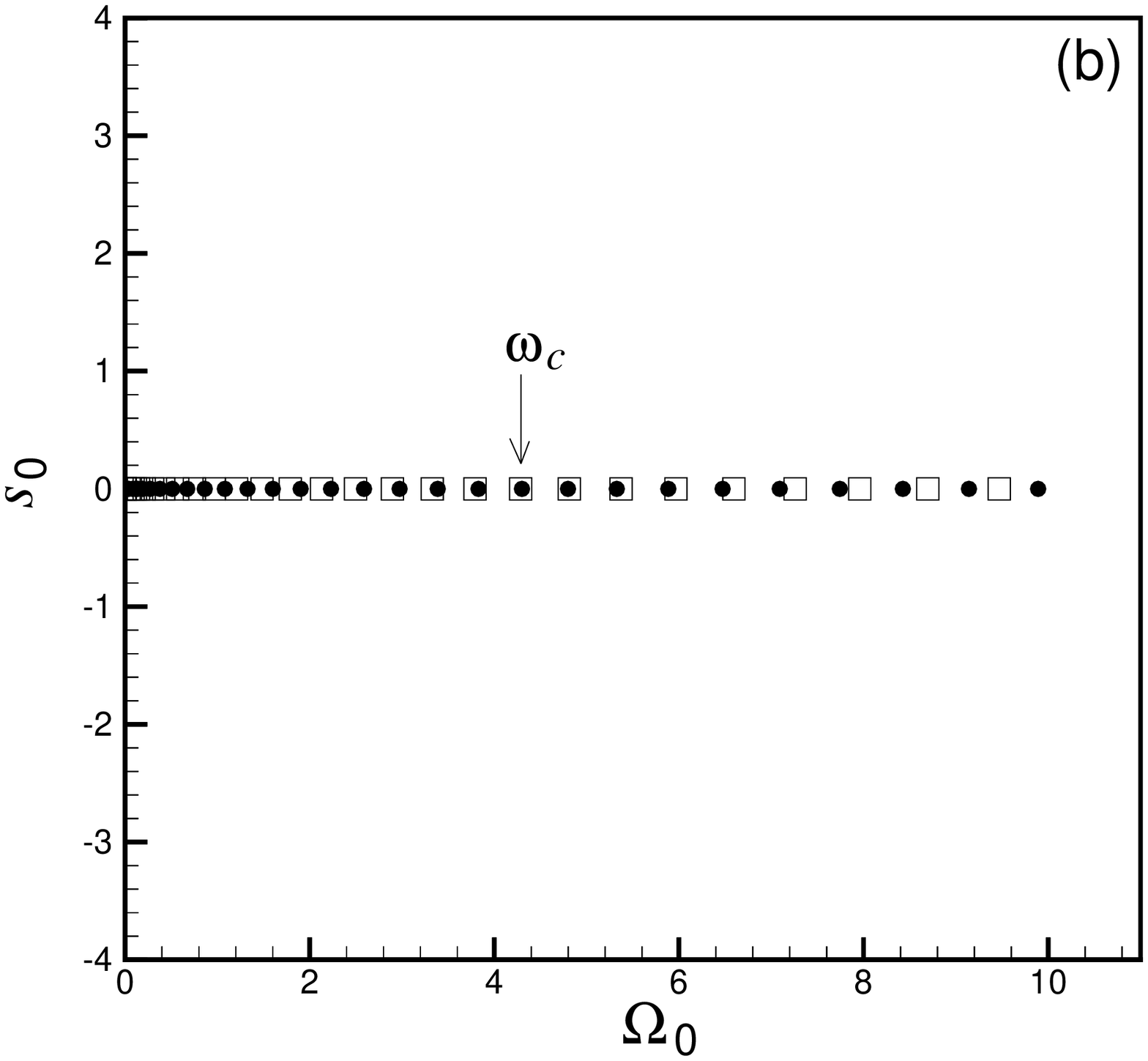}}
            \hspace{0.2truecm}
            \hbox{\epsfxsize=5.5truecm\epsfbox{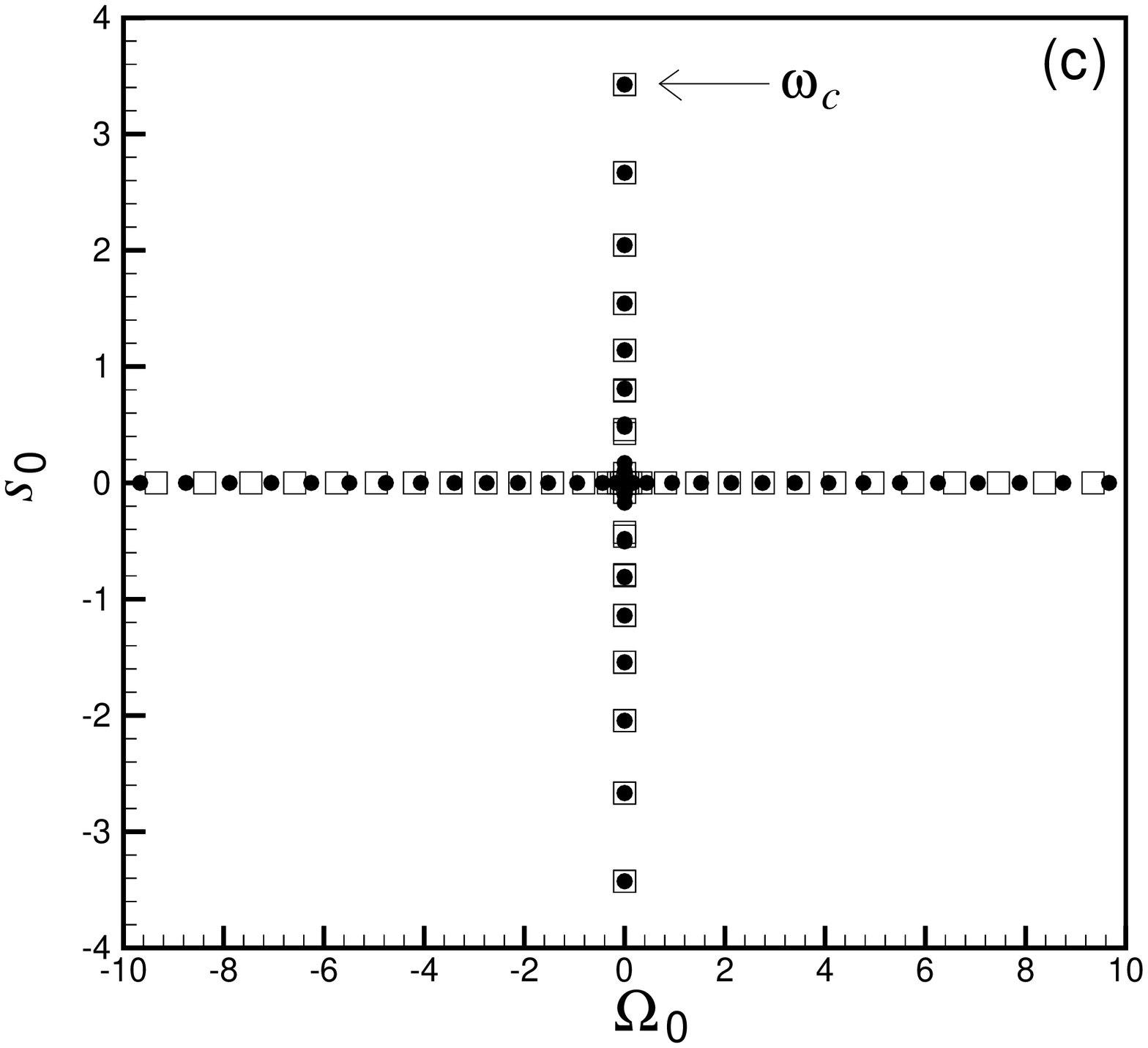}}}
\centerline{\hbox{\epsfxsize=5.5truecm\epsfbox{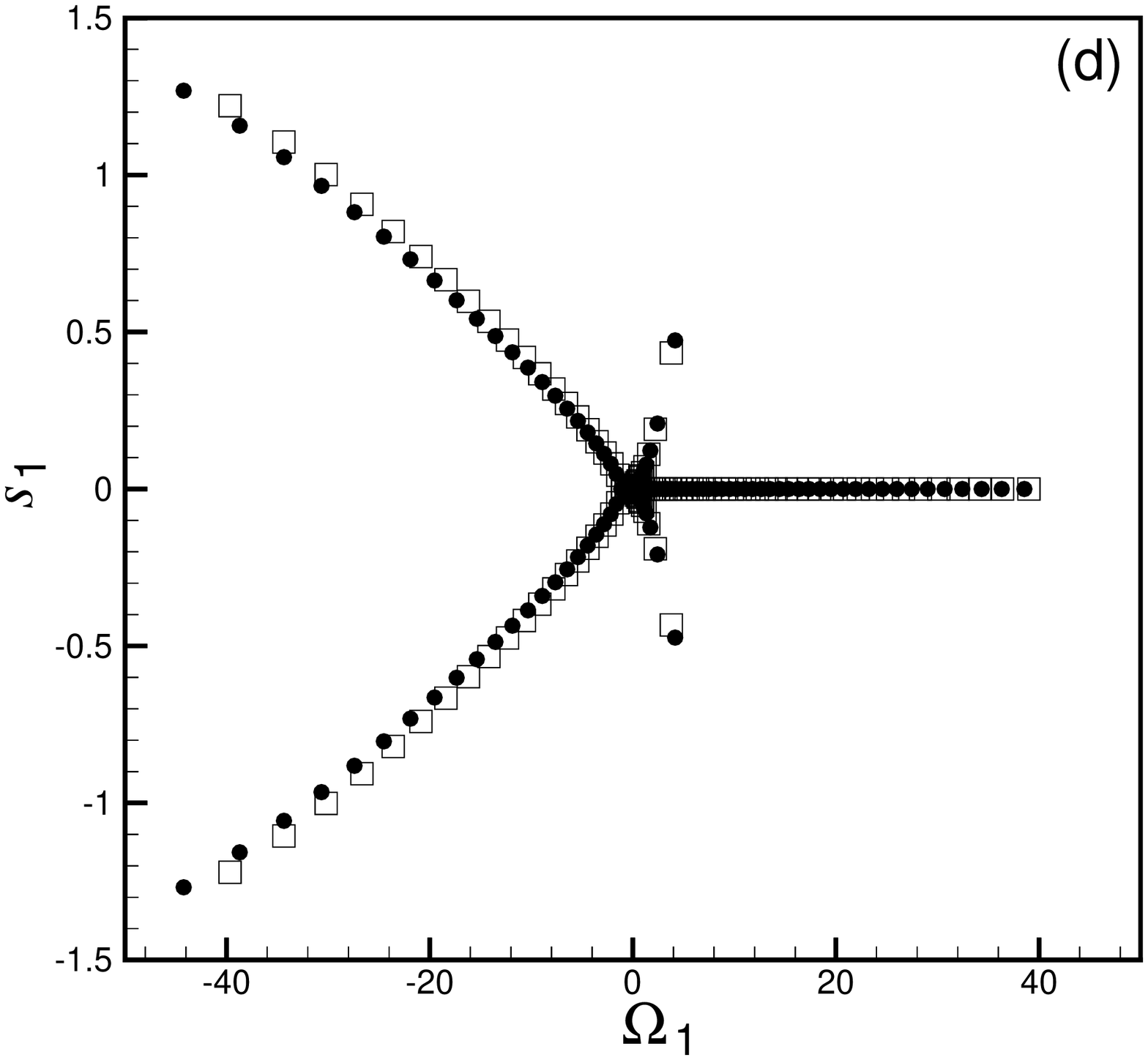}}
            \hspace{0.2truecm}
            \hbox{\epsfxsize=5.5truecm\epsfbox{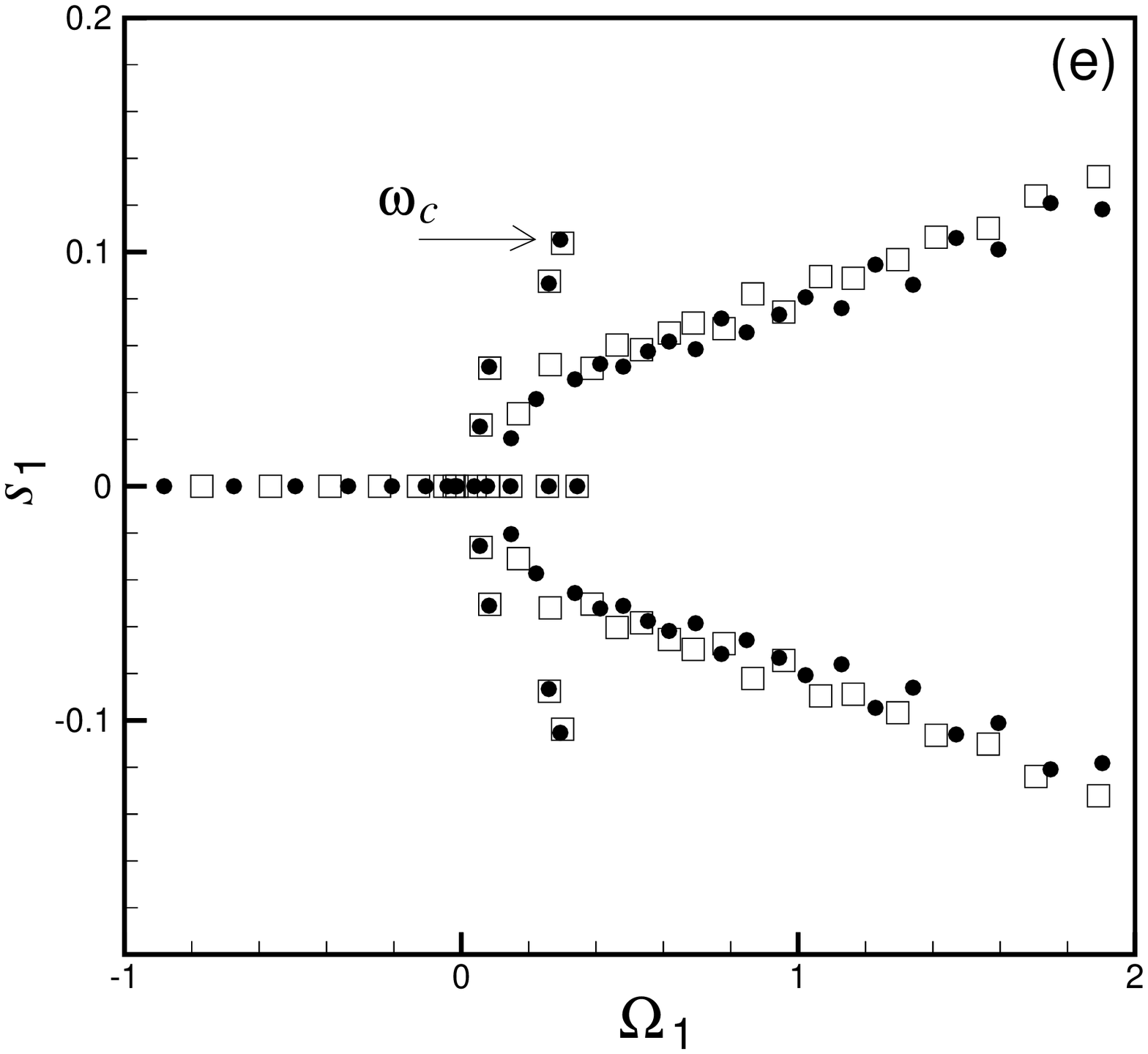}}
            \hspace{0.2truecm}
            \hbox{\epsfxsize=5.5truecm\epsfbox{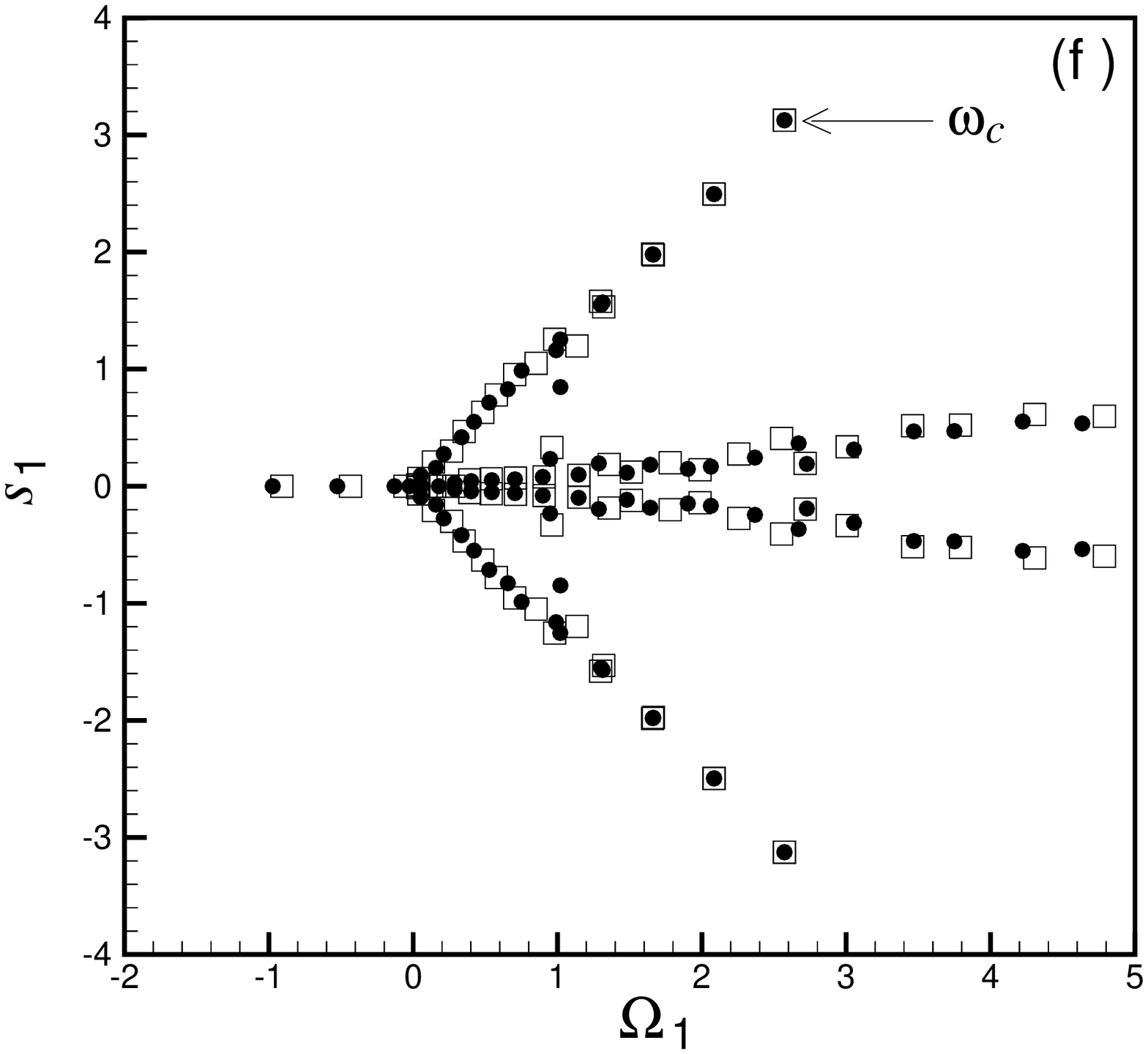}}}
\centerline{\hbox{\epsfxsize=5.5truecm\epsfbox{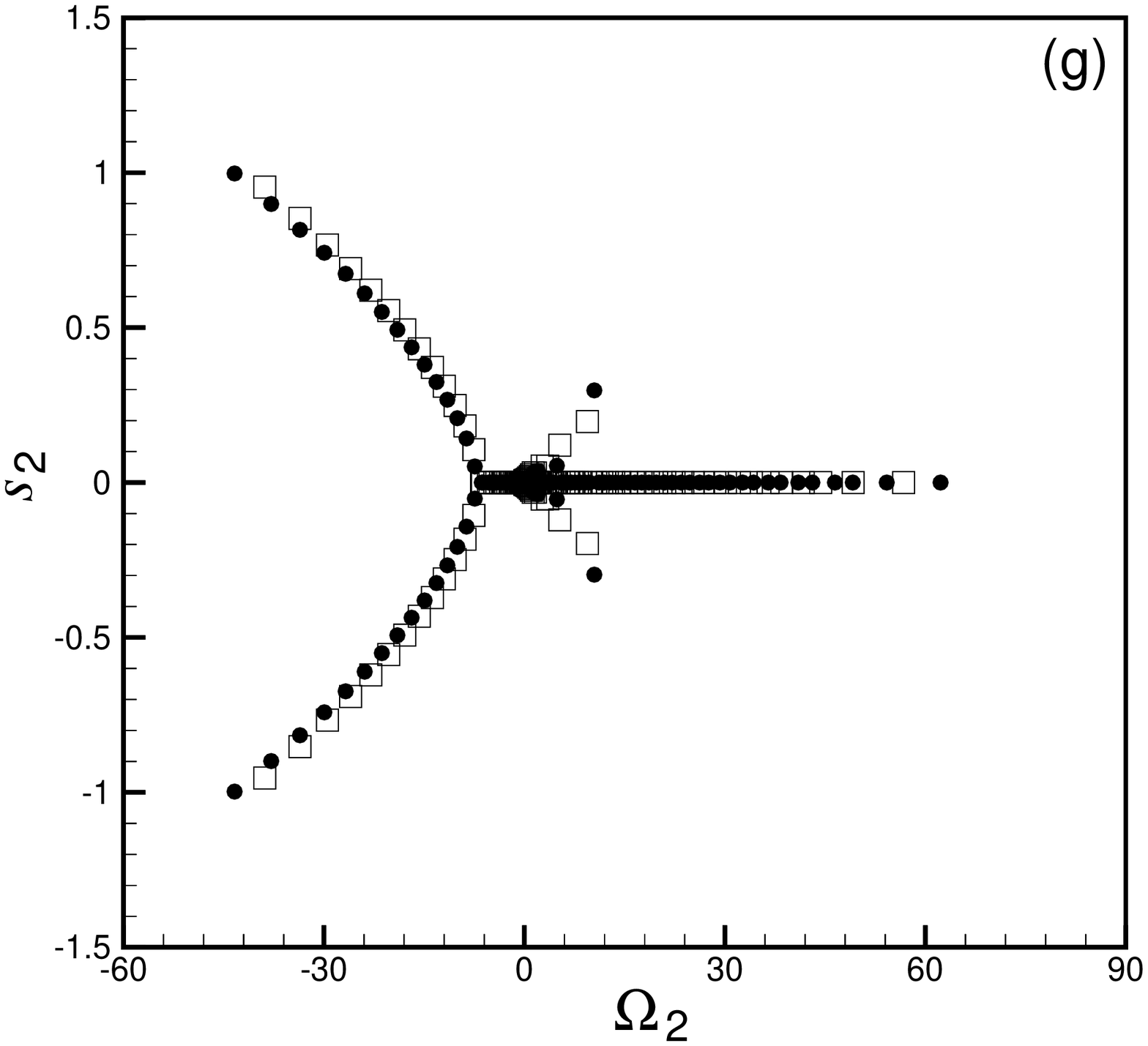}}
            \hspace{0.2truecm}
            \hbox{\epsfxsize=5.5truecm\epsfbox{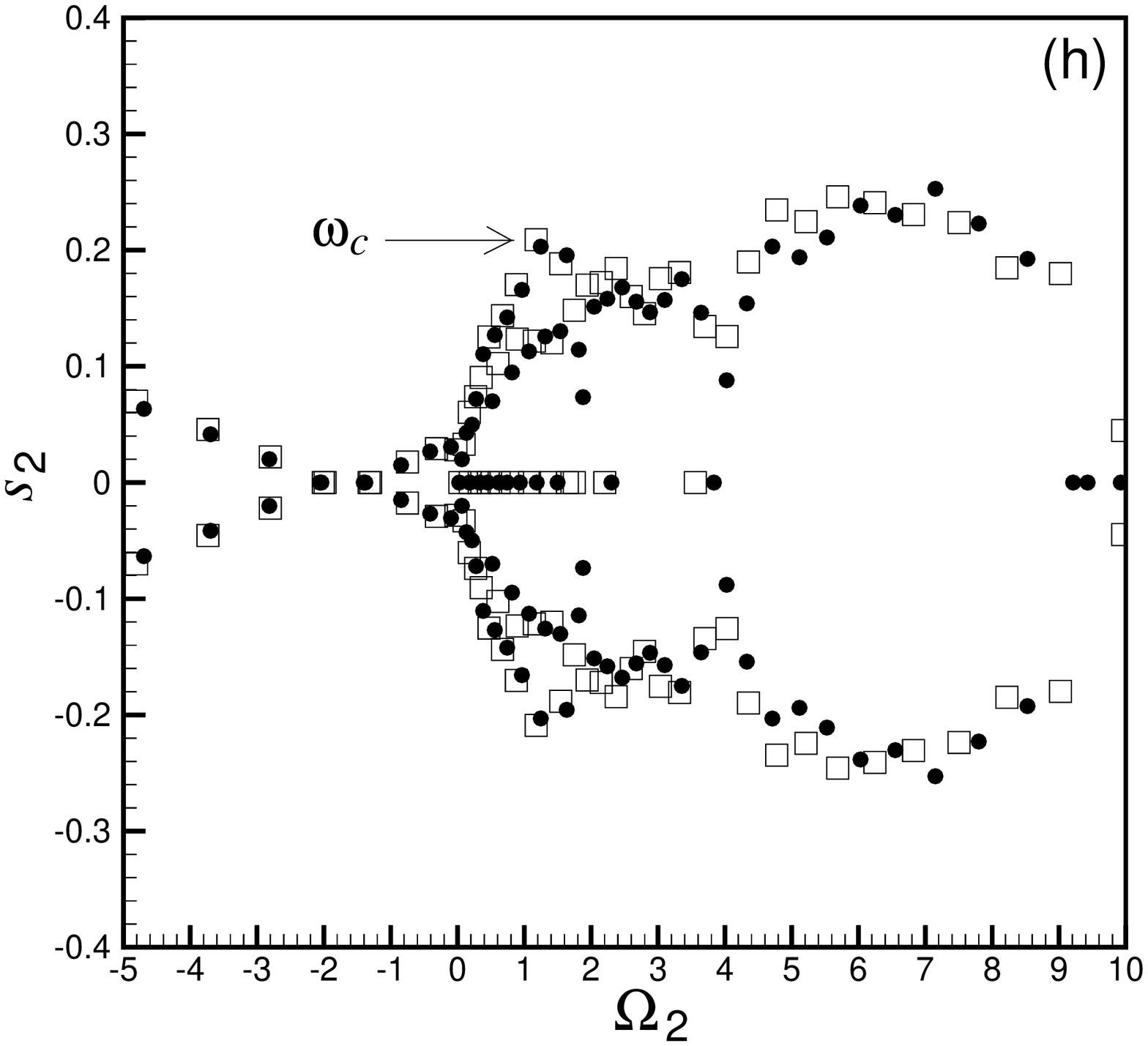}}         
            \hspace{0.2truecm}
            \hbox{\epsfxsize=5.5truecm\epsfbox{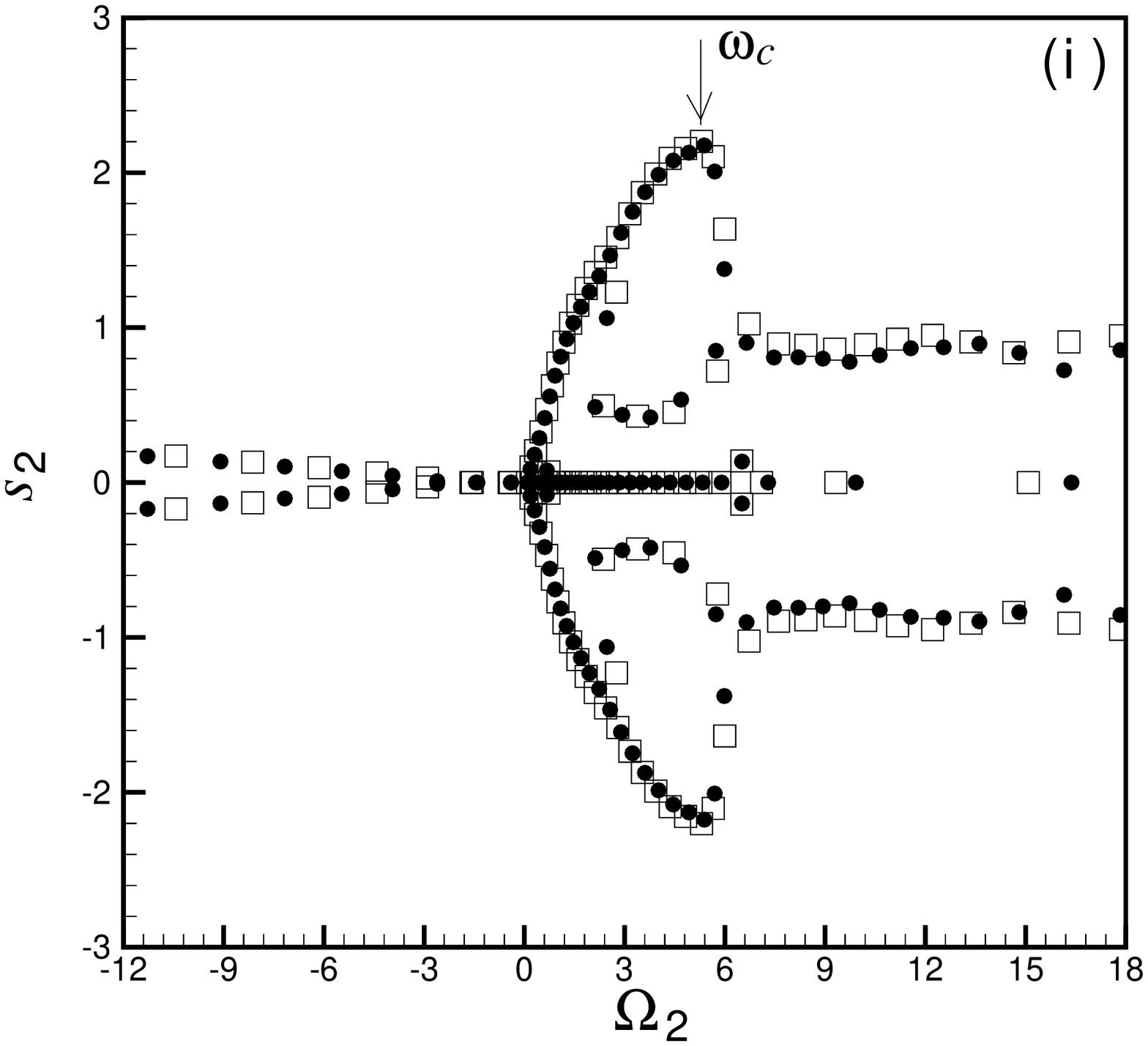}}}
\caption{Eigenvalue spectra of circular discs for $\alpha=0.5$.
Squares and filled circles correspond to $J$=50 and $J$=55,
respectively. The converged fundamental eigenfrequency has been
shown by $\omega_c$. Top, middle and bottom rows correspond to
$m=0$, 1 and 2, respectively. Left, middle and right columns are
associated with $\Mach$=0.7, 2.5 and 5, respectively.}
\label{fig:fig2}
\end{figure*}

\begin{figure*}
\centerline{\hbox{\epsfxsize=5.5truecm\epsfbox{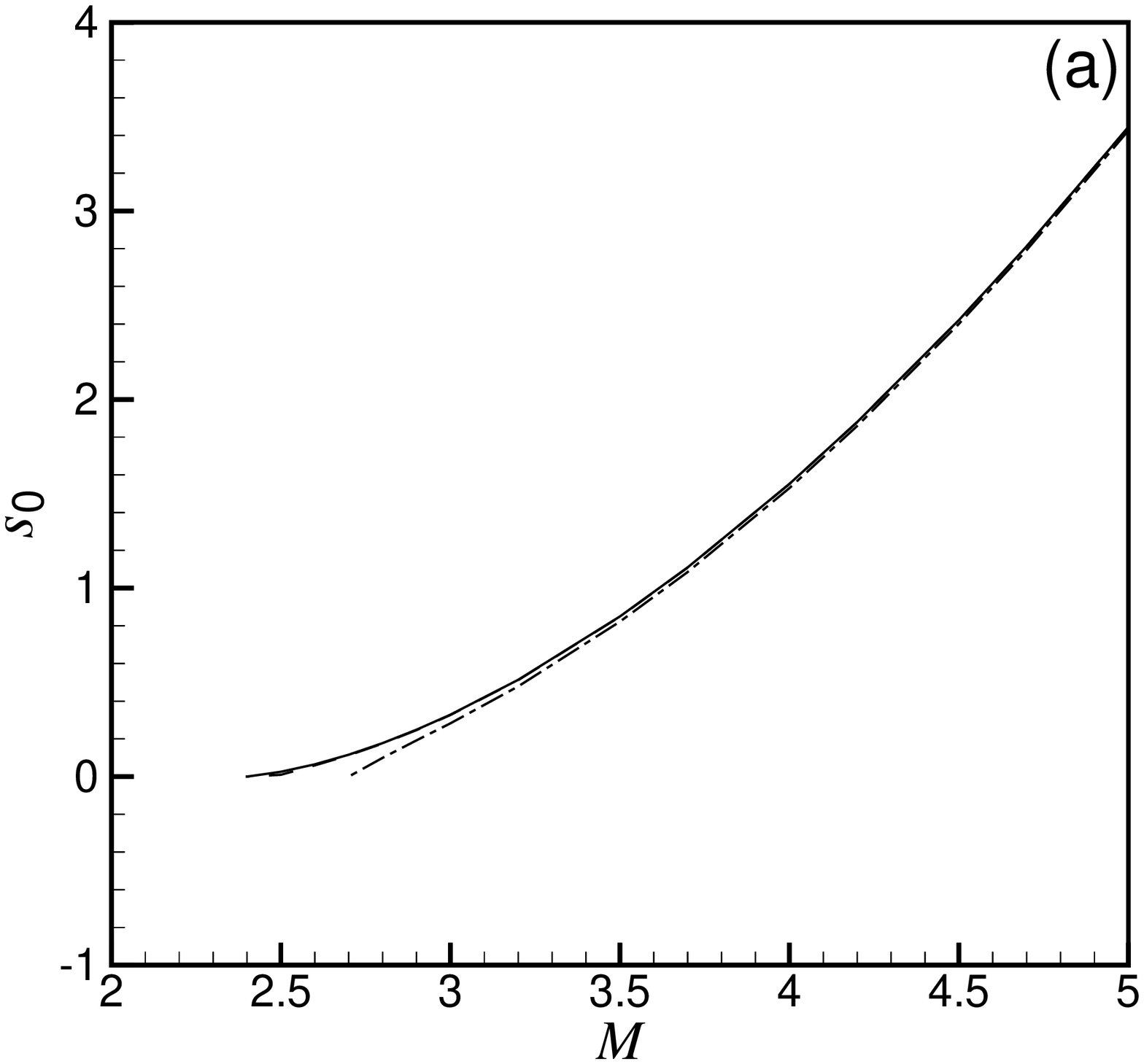}}
            \hspace{0.2truecm}
            \hbox{\epsfxsize=5.5truecm\epsfbox{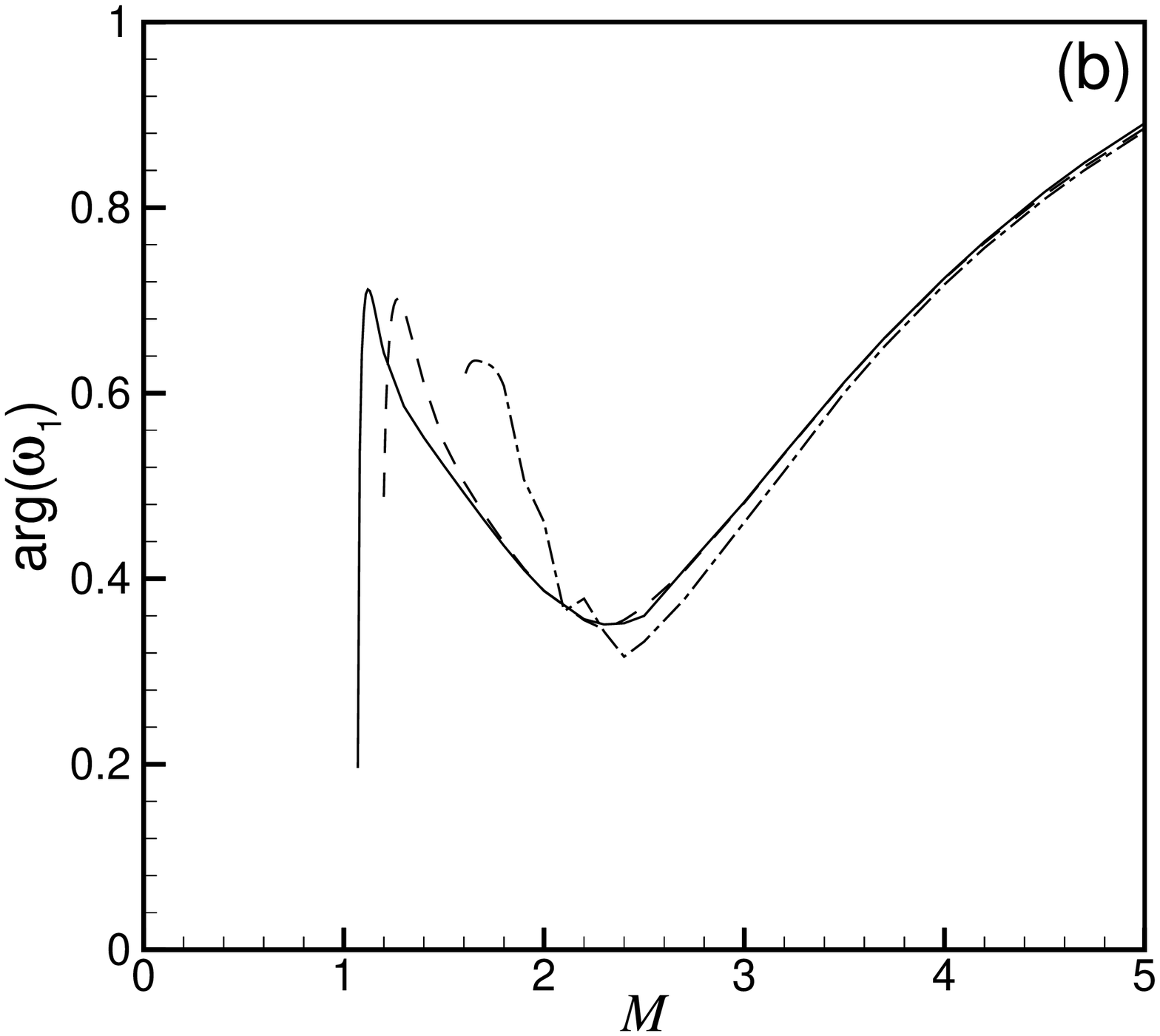}}
            \hspace{0.2truecm}
            \hbox{\epsfxsize=5.5truecm\epsfbox{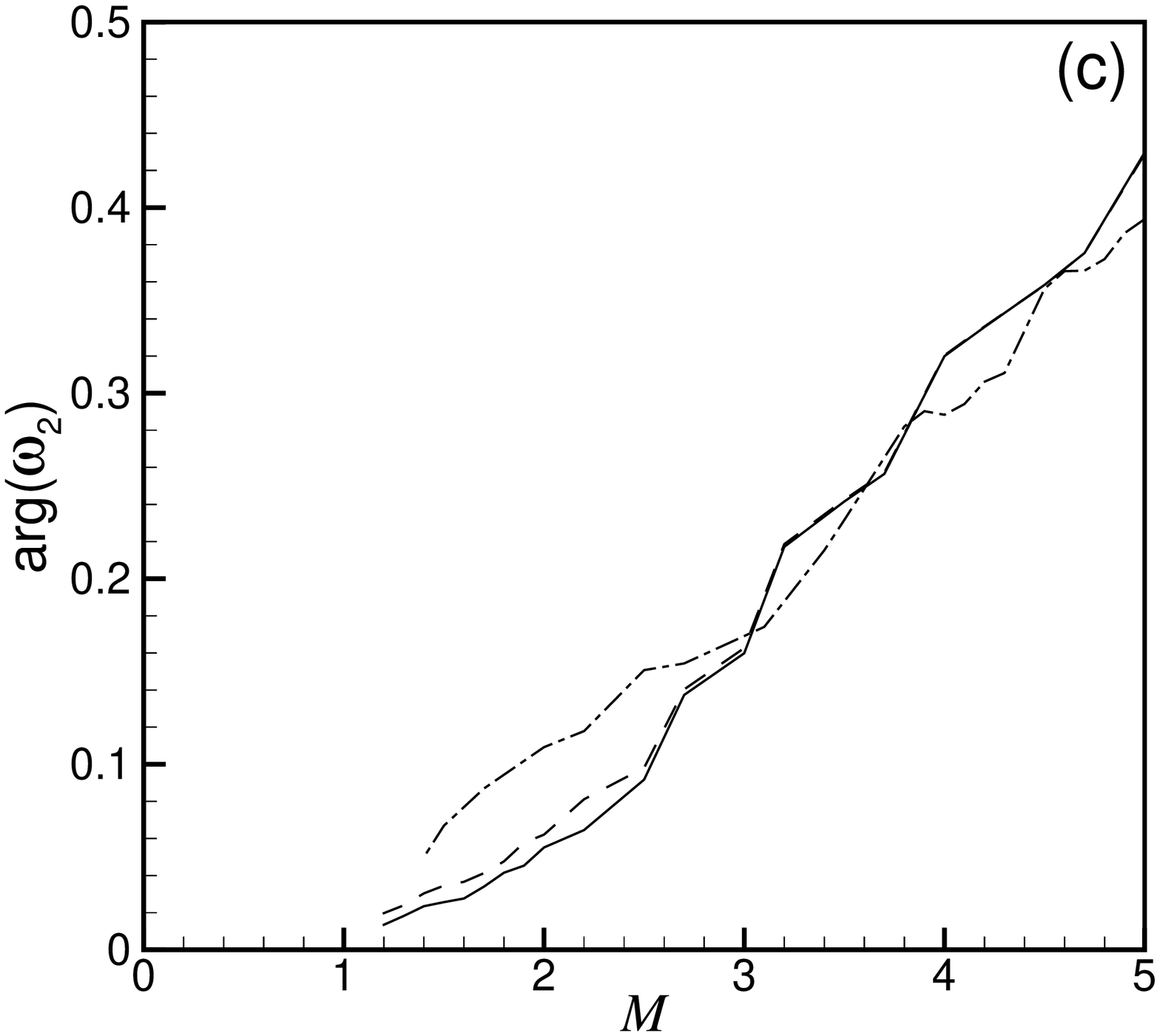}}}
\caption{Variation of the fundamental eigenvalue of circular discs
versus $\Mach$. Solid, dashed and dotted lines correspond to
$\alpha$=0.03, 0.1 and 0.5, respectively. (a) The growth rate $s_0$
of the $m=0$ mode. For this mode we have $\Omega_0=0$. (b) The
variation of ${\rm arg}(\omega_1)$ versus $\Mach$. The argument of
$\omega_1$ abruptly drops to zero at a critical $\Mach$. This
critical value tends to 1 as $\alpha$ is decreased. (c) Same as
Panel {\em b} but for $m=2$. The argument of $\omega_2$
monotonically tends to zero by decreasing $\Mach$.} \label{fig:fig3}
\end{figure*}

\subsection{Matrix formulation}
\label{sec::eigenvalue-problem} We now substitute from
(\ref{eq:bi-orthogonal-potential}) and
(\ref{eq:bi-orthogonal-u-vector}) into
(\ref{eq::perturbed-vec-m-mode}), left-multiply both sides of the
resulting equation by $-{\rm i}\hat P^{\vert
k\vert}_j\textbf{W}^{-1}$ ($k=0,1,2,\cdots$), and integrate over
$0\le R <\infty$. Using the orthogonality of the associated Legendre
functions, we obtain a system of linear equations for the
coefficients $\hat a^k_j$, $\hat b^k_j$ and $\hat c^k_j$ as
\begin{eqnarray}
\!\!\!&&\!\!\! \textbf{A}\cdot \textbf{X} = \frac{\omega_m
R_0}{c_0}\textbf{X},
\label{eq:eigenvalue-equation} \\
\!\!\!&&\!\!\! \textbf{X}^T = \left(\ldots,
\textbf{a}_{m+n},\textbf{b}_{m+n},\textbf{c}_{m+n},
\textbf{a}_{m},\textbf{b}_{m},
\textbf{c}_{m},\textbf{a}_{m-n},\ldots \right ), \nonumber \\
\!\!\!&&\!\!\! \textbf{a}_r = \left(\hat a^r_{|r|},\hat
a^r_{|r|+1},\hat a^r_{|r|+2},\ldots \right ),
\nonumber \\
\!\!\! && \!\!\! \textbf{b}_r = \left (\hat b^r_{|r|},\hat
b^r_{|r|+1},\hat b^r_{|r|+2},\ldots \right ),
\nonumber \\
\!\!\! && \!\!\! \textbf{c}_r = \left (\hat c^r_{|r|},\hat
c^r_{|r|+1},\hat c^r_{|r|+2},\ldots \right ). \nonumber
\end{eqnarray}
It is seen that $\omega_m R_0/c_0$ is an eigenvalue of $\textbf{A}$
and $\textbf{X}$ is its corresponding eigenvector. We find that the
elements of $\textbf{A}$ are linear combinations of the following
definite integrals
\begin{eqnarray}
I_{ij}^{r s}(\nu ) \!\!\!\!\! &=& \!\!\!\!\! \int_{-1}^{1} \!\!
\hat{P}^{|r|}_i\left(\xi\right) \xim^\nu \hat{P}^{|s|}_j
\left(\xi\right) d\xi,
\nonumber \\
J_{ij}^{r s}(\nu ) \!\!\!\!\! &=& \!\!\!\!\!
\int_{-1}^{1}\!\!\hat{P}^{|r|}_i \left(\xi\right)
\xip^{-\frac{1}{2}}\xim^\nu \hat{P}^{|s|}_j \left(\xi\right) d\xi,
\nonumber \\
K_{ij}^{r s}(\nu ) \!\!\!\!\! &=& \!\!\!\!\!
\int_{-1}^{1}\!\!\hat{P}^{|r|}_i\left(\xi\right)
\xip^{-\frac{1}{2}}\xim^\nu \nonumber \\
\!\!\! &{}& \!\!\!\! \times
\left\{\xim+\gamma^2\xip\right\}^{-1} \nonumber \\
\!\!\! &{}& \!\!\! \times \left\{\alpha^2\xim+
\xip\right\}^{-1}\hat{P}^{|s|}_j \left(\xi\right) d\xi,
\nonumber \\
L_{ij}^{r s}(\nu ) \!\!\!\!\! &=& \!\!\!\!\!
\int_{-1}^{1}\!\!\hat{P}^{|r|}_i\left(\xi\right)
\xip^{-\frac{1}{2}}\xim^\nu \nonumber \\
\!\!\! &{}& \!\!\!\! \times
\left\{\xim+\gamma^2\xip\right\}^{-2} \nonumber \\
\!\!\! &{}& \!\!\! \times \left\{\alpha^2\xim+
\xip\right\}^{-2}\hat{P}^{|s|}_j \left(\xi\right) d\xi,
\label{eq:integrals}
\end{eqnarray}
where $r$ and $s$ are integer numbers of the form $m+n p'$ and $m+n
p''$, respectively. The integrals in (\ref{eq:integrals}) are
evaluated using Gaussian quadratures. We transform $\textbf{A}$ to
Hessenberg form and obtain its corresponding QR transform (Press et
al. 1997). We then compute the eigenvalues and eigenvectors of the
linear system (\ref{eq:eigenvalue-equation}). Consequently, the
surface density of the $m$th mode is determined from
(\ref{eq::perturbed-fields-expansion}) and
(\ref{eq:bi-orthogonal-u-vector}) as
\begin{eqnarray}
\Sigma_{1,m}(R,\phi,t,\epsilon) \!\!\!\!\! &=& \!\!\!\!\! e^{s_m t}
\!\! A_m(R,\phi,\epsilon) \nonumber \\
\!\!\!\!\! &{}& \!\!\!\!\! \times \cos[m\phi \!-\! \Omega_m t \!+\!
\Theta_m(R,\phi,\epsilon)],
\label{eq::mode-nonaxi}\\
A_m(R,\phi,\epsilon) \!\!\!\!\! &=& \!\!\!\!\! \sqrt {
\Sigma^{(m)}_1(R,\phi,\epsilon)\bar\Sigma^{(m)}_1(R,\phi,\epsilon)},
\label{eq::amplitude-nonaxisymmetric}\\
\Theta_m(R,\phi,\epsilon) \!\!\!\!\! &=& \!\!\!\!\! \arctan \left [
-{\rm i}{\Sigma^{(m)}_1 -\bar\Sigma^{(m)}_1 \over \Sigma^{(m)}_1
+\bar\Sigma^{(m)}_1}\right ]-m\phi,
\label{eq::phase-nonaxisymmetric}
\end{eqnarray}
where a bar denotes complex conjugate. One can expand $A_m$ and
$\Theta_m$ in the Taylor series of $\epsilon$ as
\begin{equation}
Y_m(R,\phi,\epsilon) = Y_{m,0}(R)+\sum_{l=1}^{+\infty}\epsilon ^l
Y_{m,l}(R,\phi),~~Y\equiv (A,\Theta),
\label{eq::amplitude-expansion}
\end{equation}
so that $Y_{m,l}$ are $2\pi$-periodic in $\phi$. The leading zeroth
order terms of (\ref{eq::amplitude-expansion}), $Y_{m,0}(R)$,
determine the overall shape of the perturbed density, which is
usually an $m$-fold trailing spiral. The higher-order terms ($l>0$)
lead to density fluctuations along the major spiral arm(s).

\begin{figure*}
\centerline{\hbox{\epsfxsize=6.0truecm\epsfbox{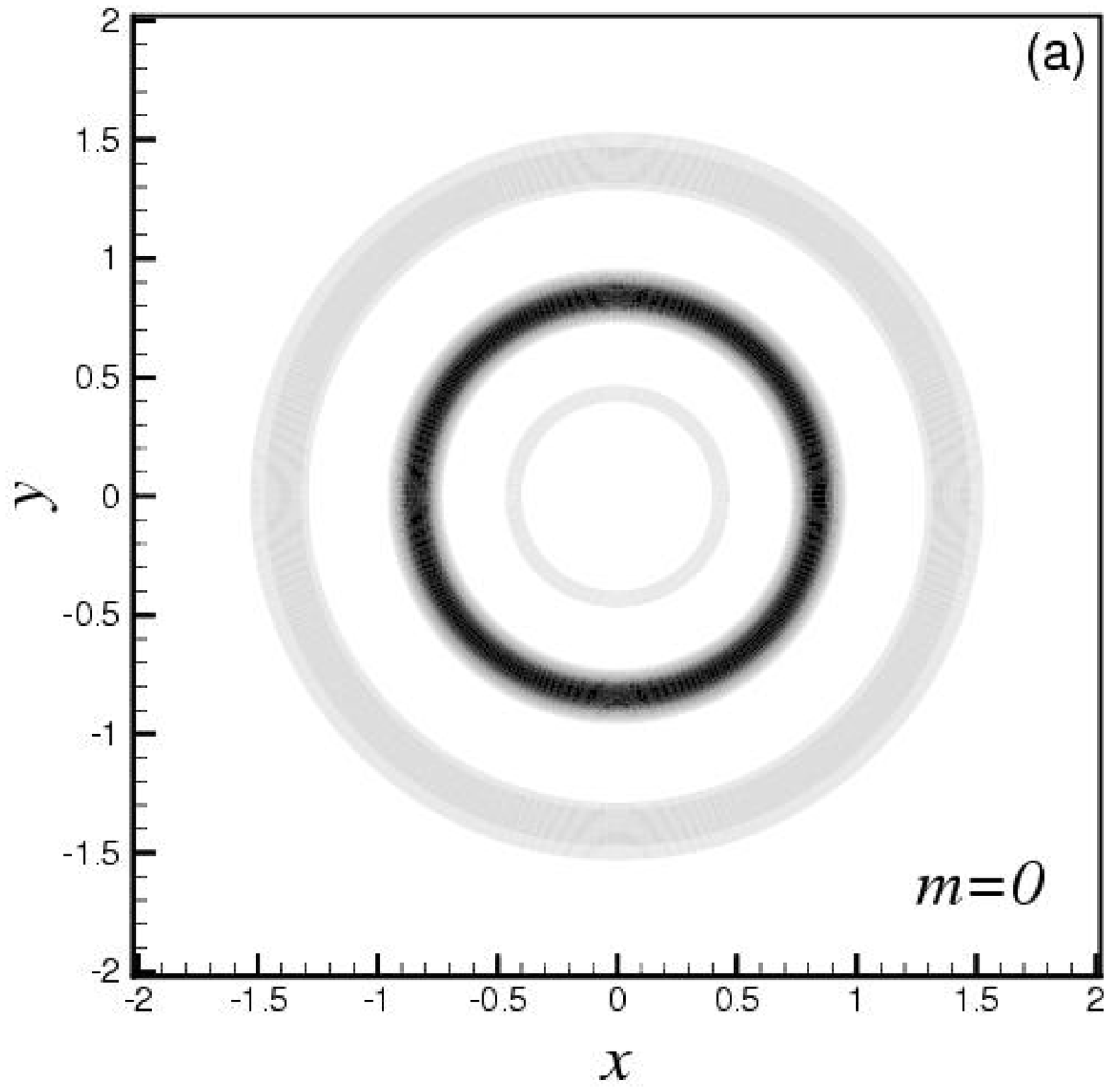}}
%            \hspace{0.2truecm}
            \hbox{\epsfxsize=6.0truecm\epsfbox{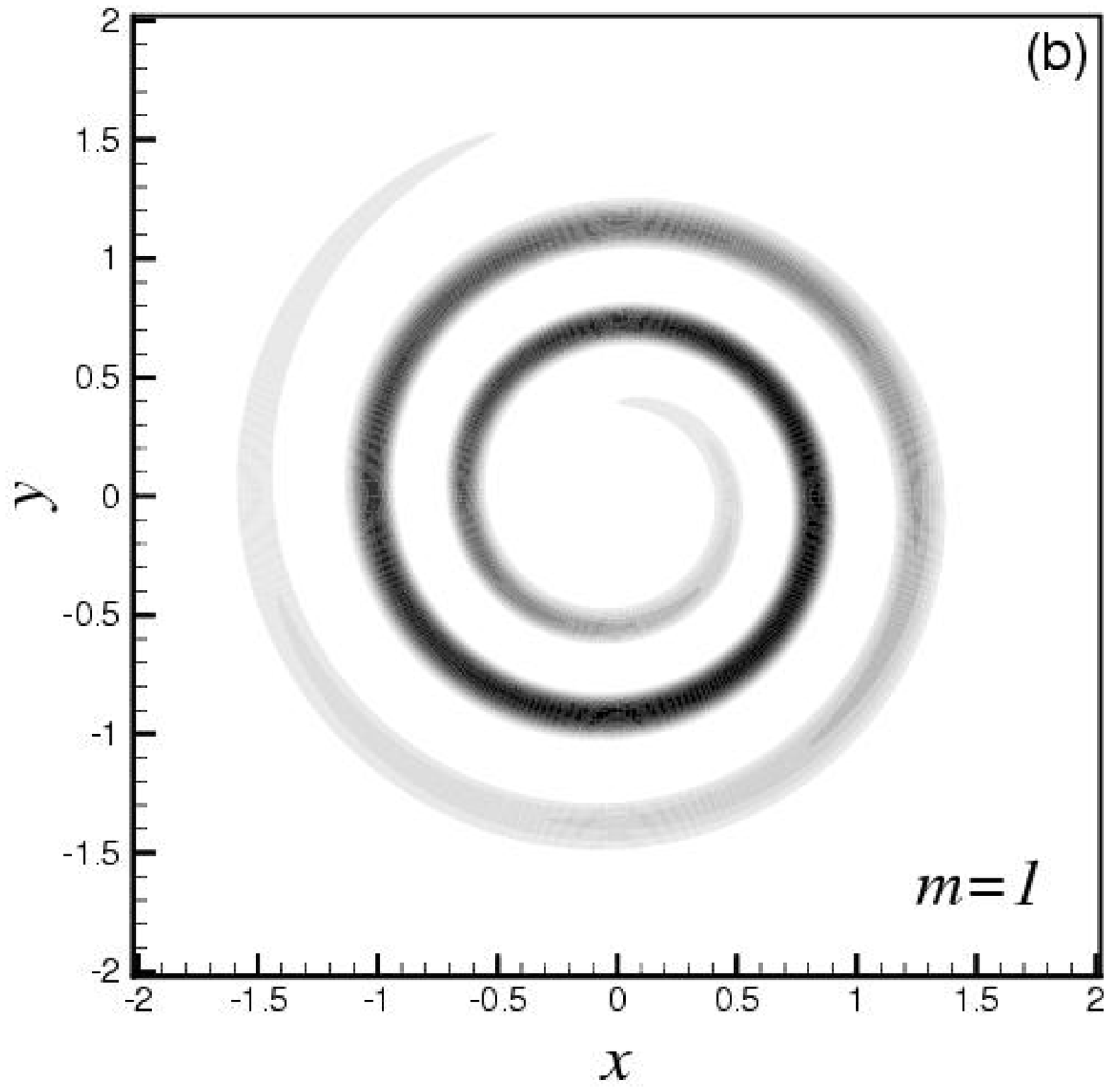}}
%            \hspace{0.2truecm}
            \hbox{\epsfxsize=6.0truecm\epsfbox{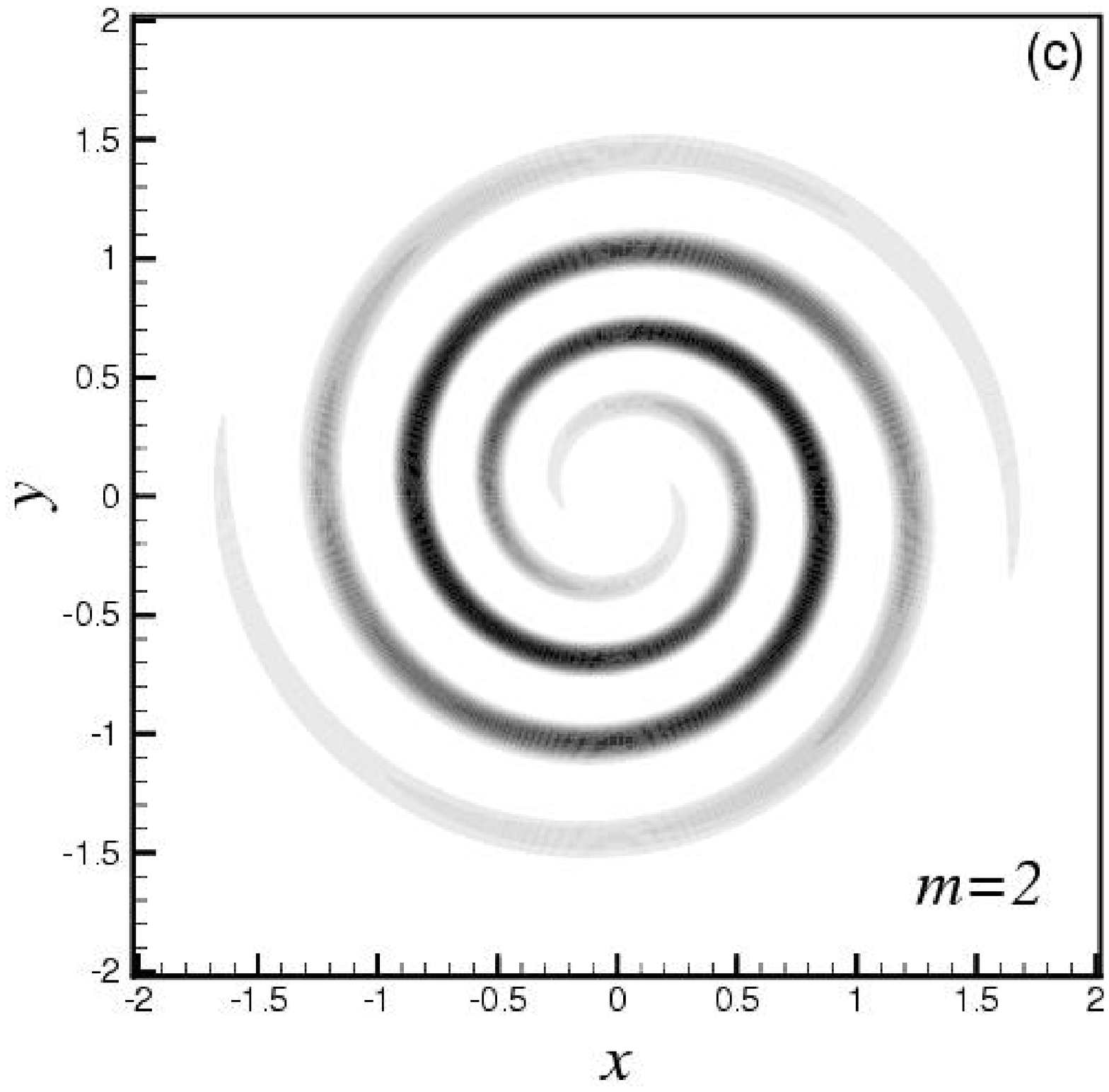}}}
\centerline{\hbox{\epsfxsize=6.0truecm\epsfbox{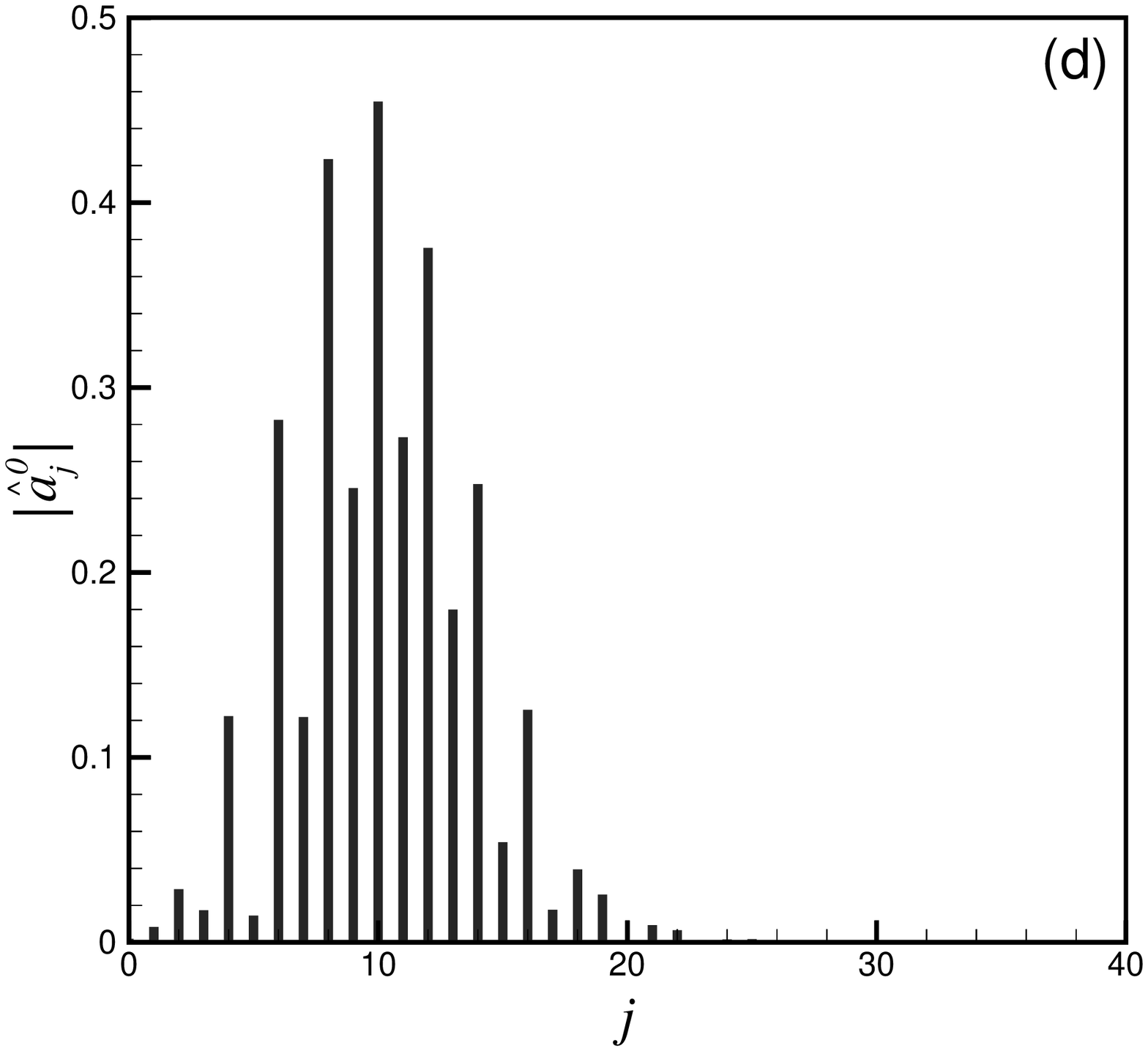}}
%            \hspace{0.2truecm}
            \hbox{\epsfxsize=6.0truecm\epsfbox{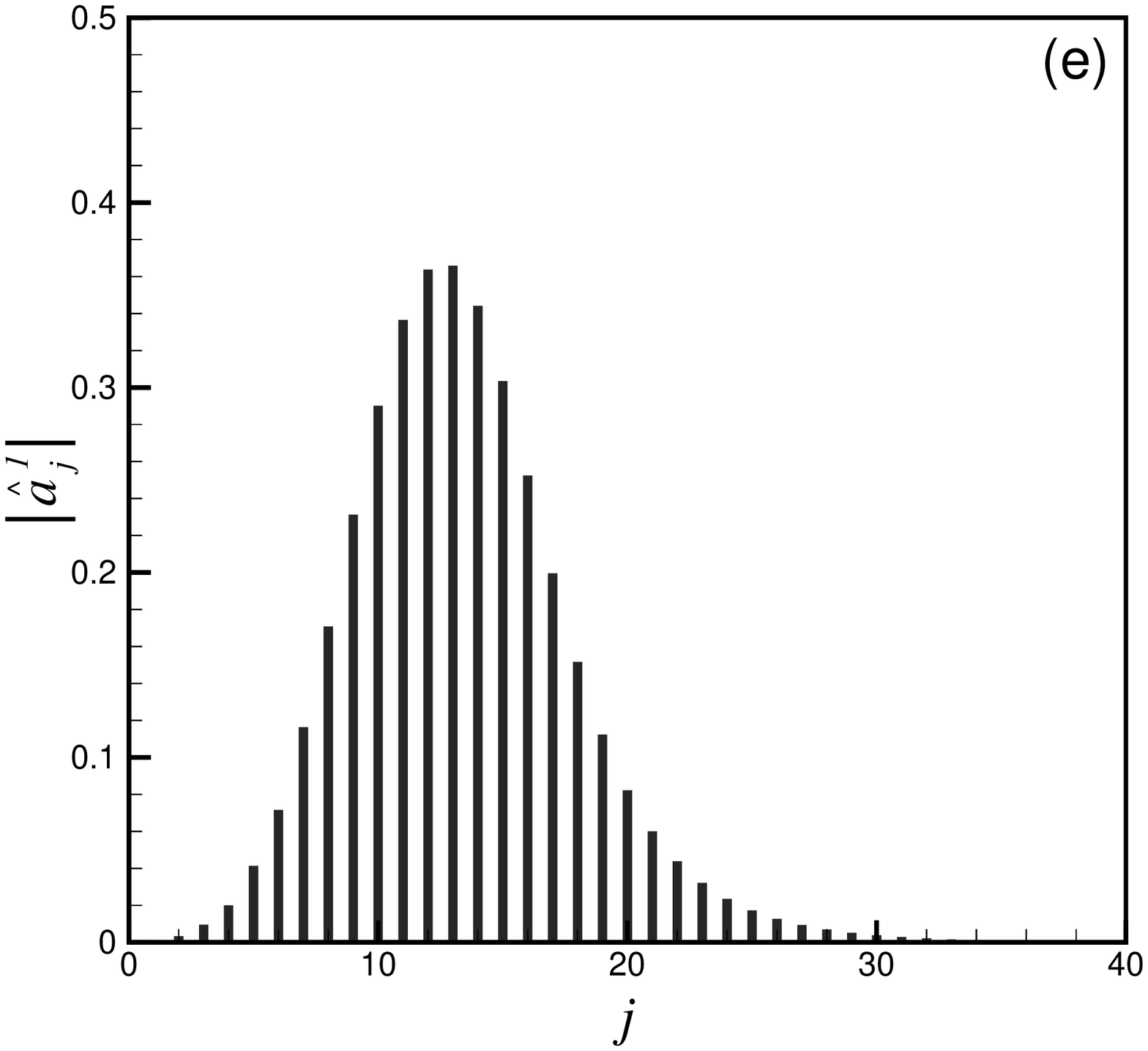}}
%            \hspace{0.2truecm}
            \hbox{\epsfxsize=6.0truecm\epsfbox{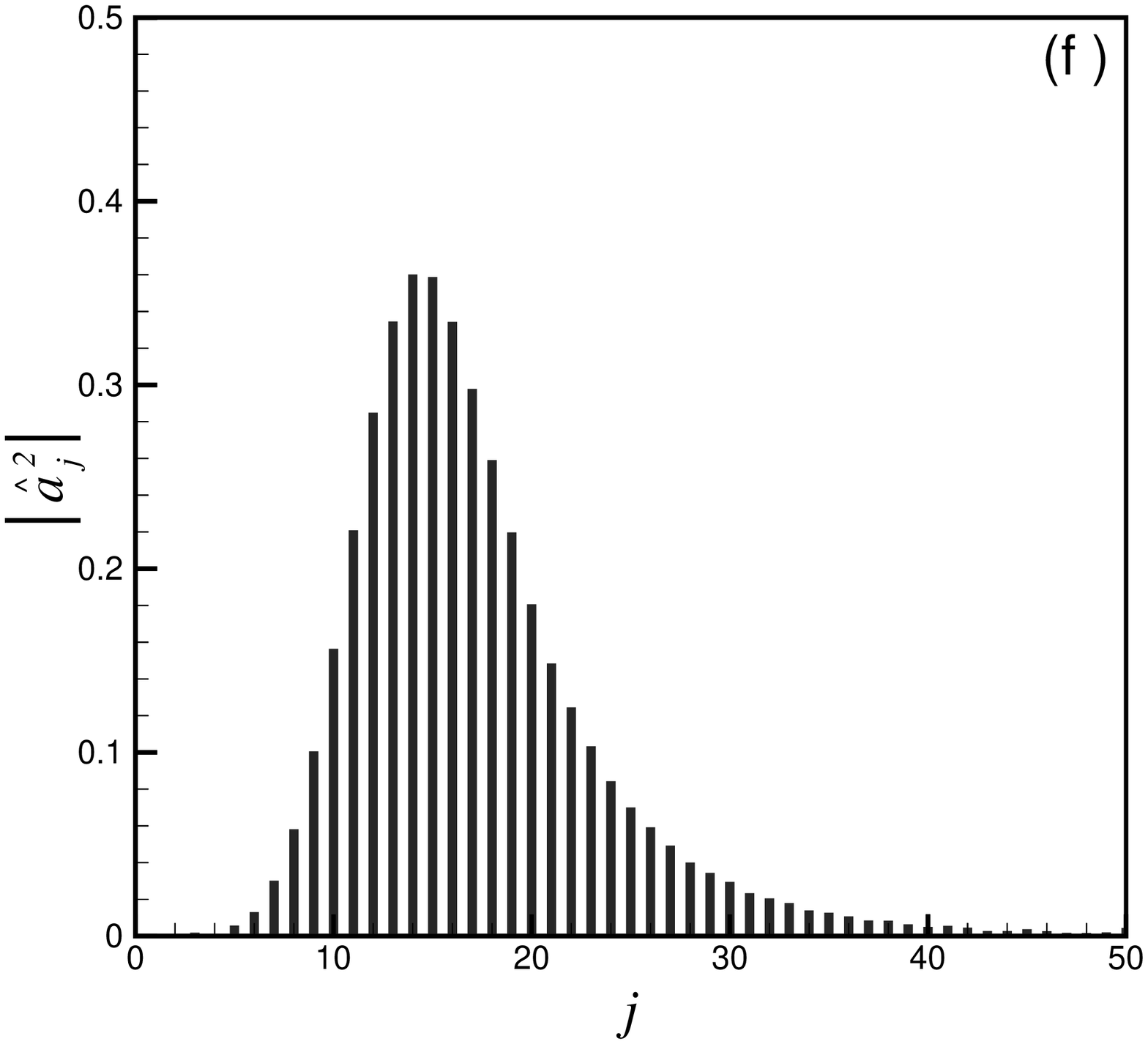}}}
\caption{{\it Top row}: unstable modes of a circular disc with
$\alpha=0.5$ and $\Mach=5$. The corresponding absolute values of
$\hat a^m_j$ ($m=0,1,2$) have been plotted versus $j$ below each
mode shape. The corotation radius is $R_{\rm CR}=0.972$ and
$R_{\rm CR}=0.943$ for $m=1$ and 2, respectively.} \label{fig:fig4}
\end{figure*}

Both $\textbf{A}$ and $\textbf{X}$ are infinite dimensional.
However, the non-axisymmetry parameter $\epsilon$ is small and its
powers, which serve as the coefficients of interacting terms, fall
off rapidly. Therefore, it is reasonable to truncate the series of
the interacting modes in (\ref{eq::perturbed-vec-m-mode}) at some
$\vert l-p\vert =L$. We also truncate the series in
(\ref{eq:bi-orthogonal-potential}) and
(\ref{eq:bi-orthogonal-u-vector}) at $j_{\rm max}=J$. We begin with
$L=1$ and $J=20$, and increase them until we acquire a relative
accuracy of 1\% in computing $\omega_m$.

In the axisymmetric limit, $\textbf{A}$ becomes a block diagonal
matrix and we obtain the following eigenvalue problem for the $m$th
mode
\begin{eqnarray}
\!\!\!\!\!\! && \!\!\! \textbf{B}(m) \cdot \textbf{Z} =
\frac{\omega_m R_0}{c_0}\textbf{Z}, 
\label{eq:axisymmetric-eigenvalue-equation} \\
\!\!\!\!\!\!&&\!\!\! \textbf{Z}^T = \left( \hat a^m_1,\ldots,\hat
a^m_J,\hat b^m_1,\ldots,\hat b^m_J,\hat c^m_1,\ldots,\hat c^m_J
\right), \nonumber
\end{eqnarray}
with $\textbf{B}(m)$ being a single $3J\times 3J$ block of
$\textbf{A}$. The perturbed surface density corresponding to
(\ref{eq:axisymmetric-eigenvalue-equation}) becomes
\begin{equation}
\Sigma _{1,m}(R,\phi,t) = e^{s_m t}A_{m,0}(R) \cos \left [
m\phi \!-\! \Omega_m t \!+\! \Theta _{m,0}(R) \right ].
\label{eq:mode-density-axisymmetric-disc}
\end{equation}
The matrix $\textbf{B}$ has $3J$ eigenvalues, which lie on several
branches in the complex $(\Omega_m,s_m)$-plane. However, eigenvalues
of most branches diverge as $J$ is increased. They are virtual roots
of the determinantal equation
\begin{equation}
\left \vert \textbf{B}(m) - \frac{\omega_m R_0}{c_0}\cdot \textbf{I}
\right \vert =0,
\end{equation}
with $\textbf{I}$ being the unit matrix. Physical eigenvalues of
unstable discs (if they exist) belong to a convergent complex
branch. The convergent eigenvalue with the largest growth rate
corresponds to the fundamental eigenmode.

The ideal scale-free limit will be attained if we let both $\alpha$
and $\gamma$ tend to zero. Our calculations show that the
stability/instability properties do not change substantially by
decreasing $\gamma$. The choice of $\gamma=0.05$ gives accurate and
robust results in most cases. So we need to play with $\alpha$.
Since the scale length of Clutton-Brock functions is fixed, by
decreasing $\alpha$ we require too many terms in the expansion of
$\hat\delta^{(k)}_1(R)$ to get the series converged. The smallest
value of $\alpha$ that we could approach, while assuring
computational accuracy, was $\alpha=0.03$.

\section{Modes of circular discs}
\label{sec::modes-of-circular-discs} To this end, we study
instabilities of cutout Mestel discs, which are obtained from the
models of \S\ref{sec:non-axi-symmetric-discs} by setting $\epsilon
=0$. Calculation of unstable modes of circular discs is
straightforward because the equilibrium fields do not depend on the
azimuthal angle $\phi$. We set $m=0$, 1 and 2 in
(\ref{eq:axisymmetric-eigenvalue-equation}) and calculate $\omega_m$
over the interval $0 \le \Mach \le 5$. This covers all subsonic
(hot) and a wide range of supersonic (cold) discs. We calculate the
normal modes for $\alpha$=0.5, 0.1 and 0.03, which respectively
require $J=50$, 100 and 200 for a computational accuracy of 1\%.

Top row in Fig. \ref{fig:fig2} shows the eigenfrequency spectrum of
the $m=0$ mode for $\Mach$=0.7, 2.5, 5 and $\alpha=0.5$. Most
eigenfrequencies of the model with $\Mach$=0.7 are real (Fig.
\ref{fig:fig2}{\em a}). There are a few divergent complex
eigenfrequencies whose magnitudes are increased by increasing $J$.
The complex eigenfrequencies completely disappear for $1.8 \lta
\Mach \lta 2.7$ (Fig. \ref{fig:fig2}{\em b}). A convergent complex
branch occurs for $M > 2.7$ whose eigenfrequencies correspond to
non-rotating unstable ring modes with $\Omega_m=0$ (Fig.
\ref{fig:fig2}{\em c}). The {\it fundamental} eigenfrequency,
denoted by $\omega_c$, is located at the end of the complex branch
of the spectrum. Fig. \ref{fig:fig3}{\em a} shows the variation of
the fundamental growth rate with respect to $M$ and $\alpha$. It is
seen that for $M>3$ our results do not vary that much as we change
$\alpha$. The upper limit of $M$ in stable discs is $M_{\rm cr}=
2.7$ for $\alpha=0.5$ and $M_{\rm cr}=2.43$ for $\alpha=0.03$. Our
results for $M_{\rm cr}$ can be compared with the critical
$v/\sigma$ computed from equation (6.2) of Lemos et al. (1991).
Their $v$ and $\sigma$ for the $\beta=0$ models are our $V_c$ and
$c_0$, respectively. By keeping the minus sign before the square
root in equation (6.2) of Lemos et al. (1991) (this is to guarantee
ring formation), and setting $L=1$, $\beta=0$ and $\gamma=1$, one
finds $v/\sigma=2.414$. The difference between this critical number
and $M_{\rm cr}=2.43$ of our $\alpha=0.03$ model is only 0.66\%,
which validates the accuracy of our matrix formulation based on the
expansions introduced in equations
(\ref{eq:bi-orthogonal-potential}) and
(\ref{eq:bi-orthogonal-u-vector}). We note that the choice of
$\alpha=0.03$ is quite enough for understanding the behavior in the
scale-free limit. We have also found that unstable breathing modes
do exist in subsonic discs with $\alpha=0.5$ if $\Mach < 0.56$. For
$\alpha=0.03$ we find $\Mach < 0.87$, which coincides with the upper
limit of $v/\sigma$ calculated in Lemos et al. (1991) for the
existence of breathing modes.

The eigenfrequency spectra of the $m=1,2$ modes have three branches
in the complex $(\Omega_m,s_m)$-plane. Subsonic discs have no
convergent unstable (complex) branch. They are stable under $m=1,2$
excitations (Figs \ref{fig:fig2}{\em d} and {\em g}). A continuum of
convergent real eigenfrequencies exists on the $\Omega_m$-axis,
which survives for almost all $\Mach$, although it may be shrunk to
a small interval of positive pattern speeds. Aoki et al. (1979) have
encountered similar eigenfrequencies in studying the gaseous Kuzmin
disc. Convergent real eigenfrequencies correspond to pure
oscillatory waves. Here we are interested in growing modes and we
will overlook the real branch of the spectrum. As we increase
$\Mach$, a convergent unstable branch, with the largest ${\rm
arg}(\omega_m)$=$\arctan (s_m/\Omega_m)$, occurs in the spectrum.
One can readily identify the fundamental eigenfrequency $\omega_c$
on this newly born branch. In a sufficiently cold disc, the
convergent branch attracts more eigenfrequencies of the linear
system (\ref{eq:axisymmetric-eigenvalue-equation}) and ${\rm
Im}(\omega_c)$ becomes the global maximum (see Figs
\ref{fig:fig2}{\em f} and \ref{fig:fig2}{\em i}). In Figs
\ref{fig:fig3}{\em b} and {\em c} we have illustrated ${\rm
arg}(\omega_m)$=$\arctan(s_m/\Omega_m)$ ($m=1,2$) in terms of $M$
and for several choices of $\alpha$. These figures show that the
growth rate tends to zero faster than the pattern speed while the
disc is stabilized. Instabilities are suppressed in discs with
$\Mach <1$ because pressure waves travel faster than gravitationally
generated waves. Figs \ref{fig:fig3}{\em b} and {\em c} can be
compared with Fig. 3 in GE. The graphs of ${\rm arg}(\omega_m)$ have
similar trends in our and their papers: For $m=1$, ${\rm
arg}(\omega_m)$ has a local minimum, while it is an almost monotonic
function of $\Mach$ for cold discs and for $m=2$. The major
discrepancy between our and GE results occurs when $\Mach
\rightarrow 1$. GE predicted a rapid growth of ${\rm arg}(\omega_m)$
contrary to the abrupt drop in our calculations. We ascribe the
existing discrepancies to the type of imposed internal boundary
conditions. We use a cutout, which disables the medium needed for
the propagation of incoming waves. GE, however, fix ${\rm
arg}(\omega)$ to make a reflective boundary at $R=0$.

Fig. \ref{fig:fig4} displays the unstable $m=0$, 1 and 2 modes of a
circular disc with $\Mach=5$ and $\alpha=0.5$. We have plotted the
positive part of $A_{m,0}(R)\cos[m\phi+\Theta_{m,0}(R)]$. Gray scale
contours mark 10 to 100 percent of the maximum perturbed density.
The $m=1$ and $m=2$ modes are tightly wound spirals whose amplitude
functions, $A_{1,0}(R)$ and $A_{2,0}(R)$, have only one peak. There
is no local density concentration along the spiral arms and the wave
amplitude slowly decays well within the corotation radius. We have
also plotted $\vert \hat a^m_j\vert$ versus $j$ below each mode
shape in Fig. \ref{fig:fig4}. The plots show a rapid convergence of
the sequence $\vert \hat a^m_j\vert$. As it can be seen in mode
shapes, unstable waves cannot penetrate into the centre. This is due
to the inner cutout, which damps the impinging waves by immobilizing
the mass of central regions.

\section{Modes of non-axisymmetric discs}
\label{sec:modes-nonaxi-discs} We now apply the formulation of
\S\ref{sec:perturbation-formulation} and calculate the $m=0$, 1 and
2 modes of our non-axisymmetric discs. We set $L=3$, which keeps
$nL+1$=7 interacting modes. Such a truncation has enough accuracy
for the models with $\Mach >1$, which mostly have $\epsilon <0.24$
(see Fig. \ref{fig:fig1}). The major limiting factor in the normal
mode calculation of non-axisymmetric discs is the choice of $\alpha$
whose small values need too many terms in
(\ref{eq:bi-orthogonal-potential}) and
(\ref{eq:bi-orthogonal-u-vector}) to ensure the convergency. Taking
into account the fact that the size of $\textbf{A}$ in
non-axisymmetric discs increases by a factor of $(nL+1)\times
(nL+1)$ compared to the axisymmetric limit, the run time of our
eigenvalue and eigenvector solver rises drastically. To optimize the
cost of numerical computations, we use $\alpha=0.5$ in our
non-axisymmetric discs because it requires only $J=50$ to achieve a
relative accuracy of 1\%. This is a legitimate choice because
according to the graphs of Fig. \ref{fig:fig3} the qualitative
features of normal modes do not change by varying $\alpha$.

\begin{figure*}
\centerline{\hbox{\epsfxsize=8.0truecm\epsfbox{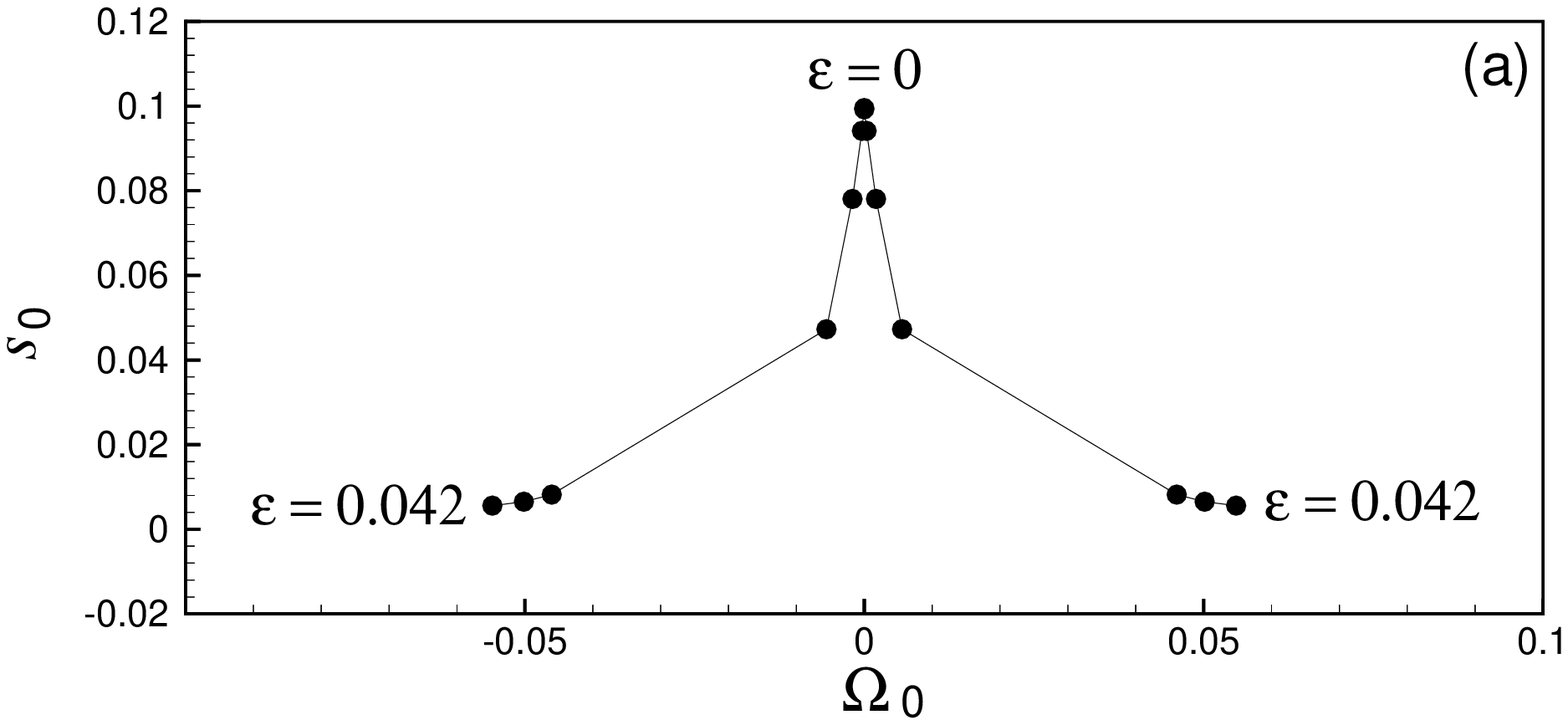}}
                  \hspace{0.3truecm}
            \hbox{\epsfxsize=8.0truecm\epsfbox{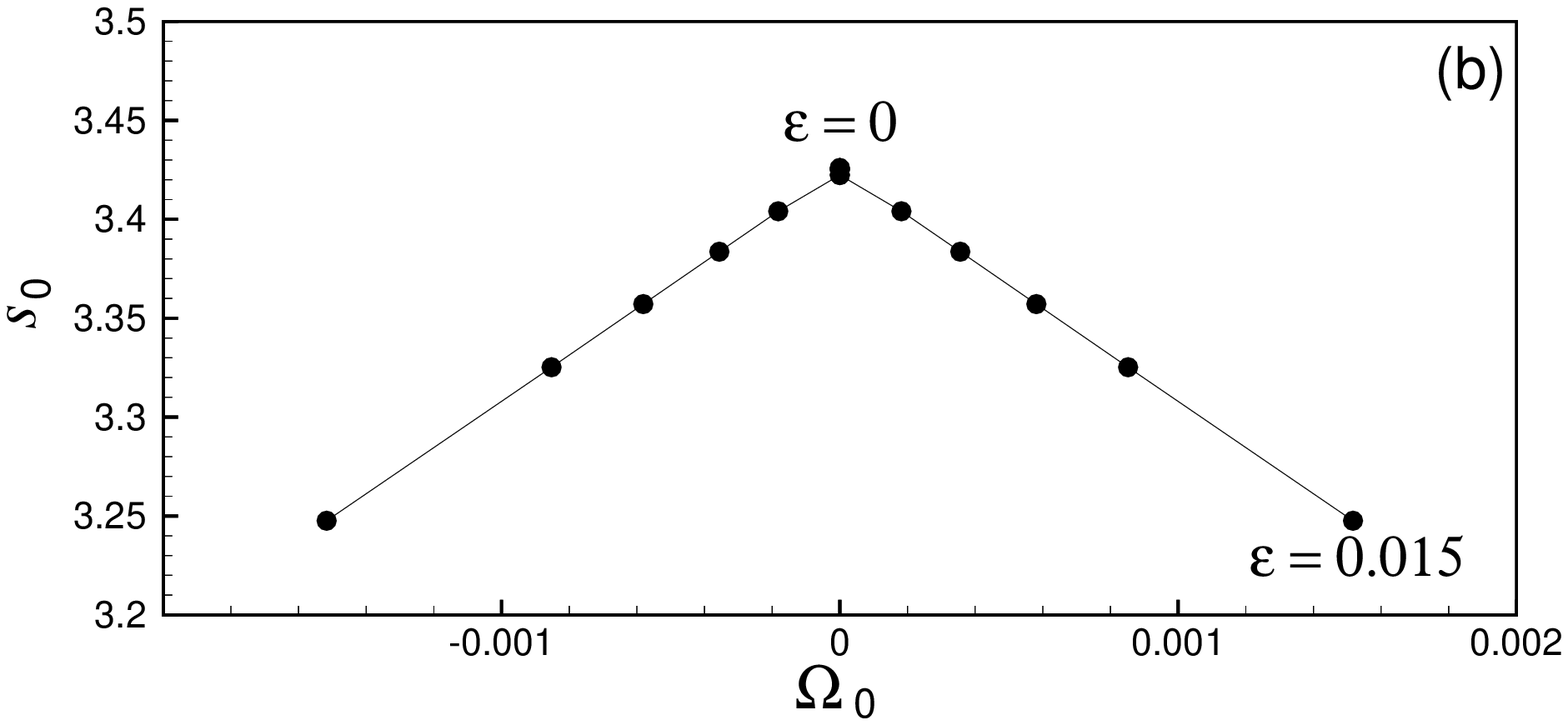}}}
\centerline{\hbox{\epsfxsize=8.0truecm\epsfbox{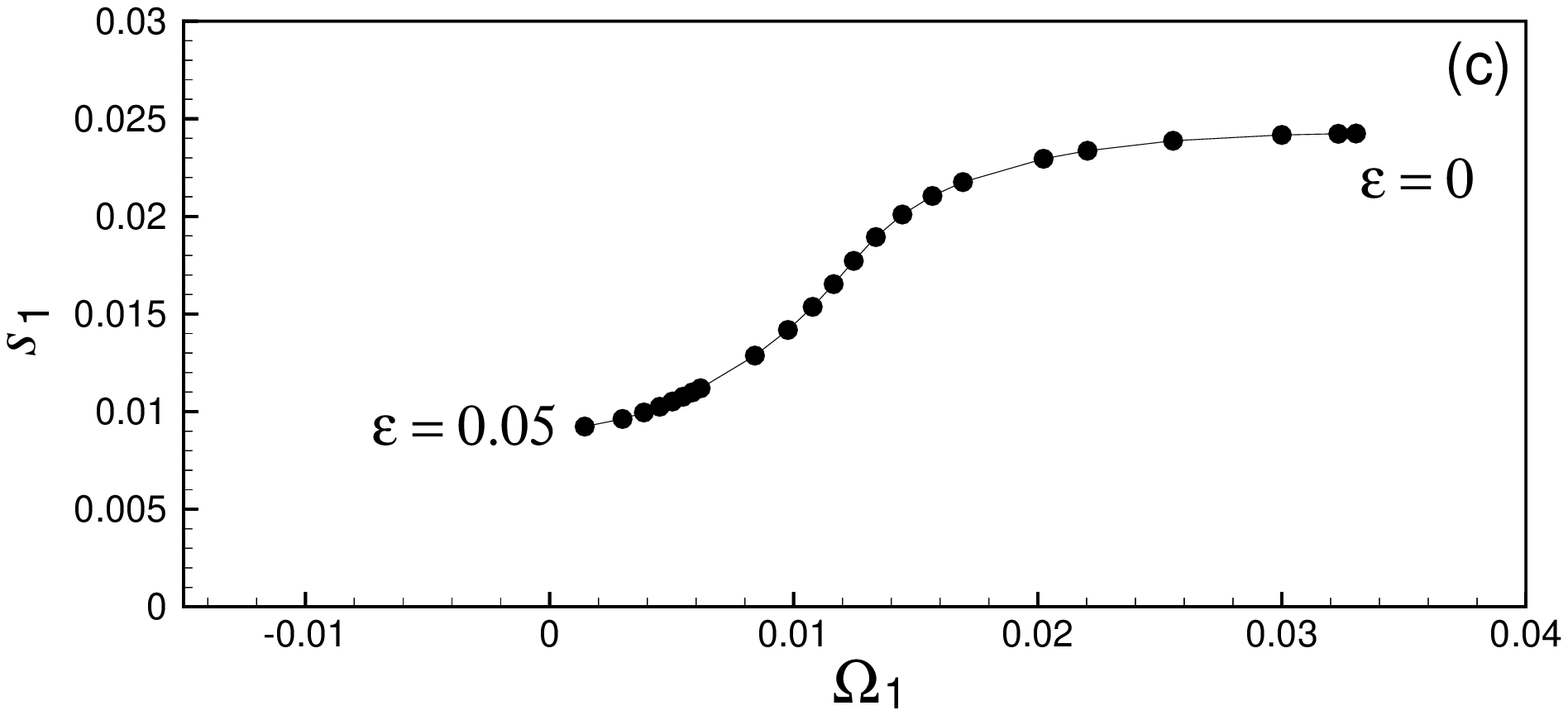}}
            \hspace{0.3truecm}
            \hbox{\epsfxsize=8.0truecm\epsfbox{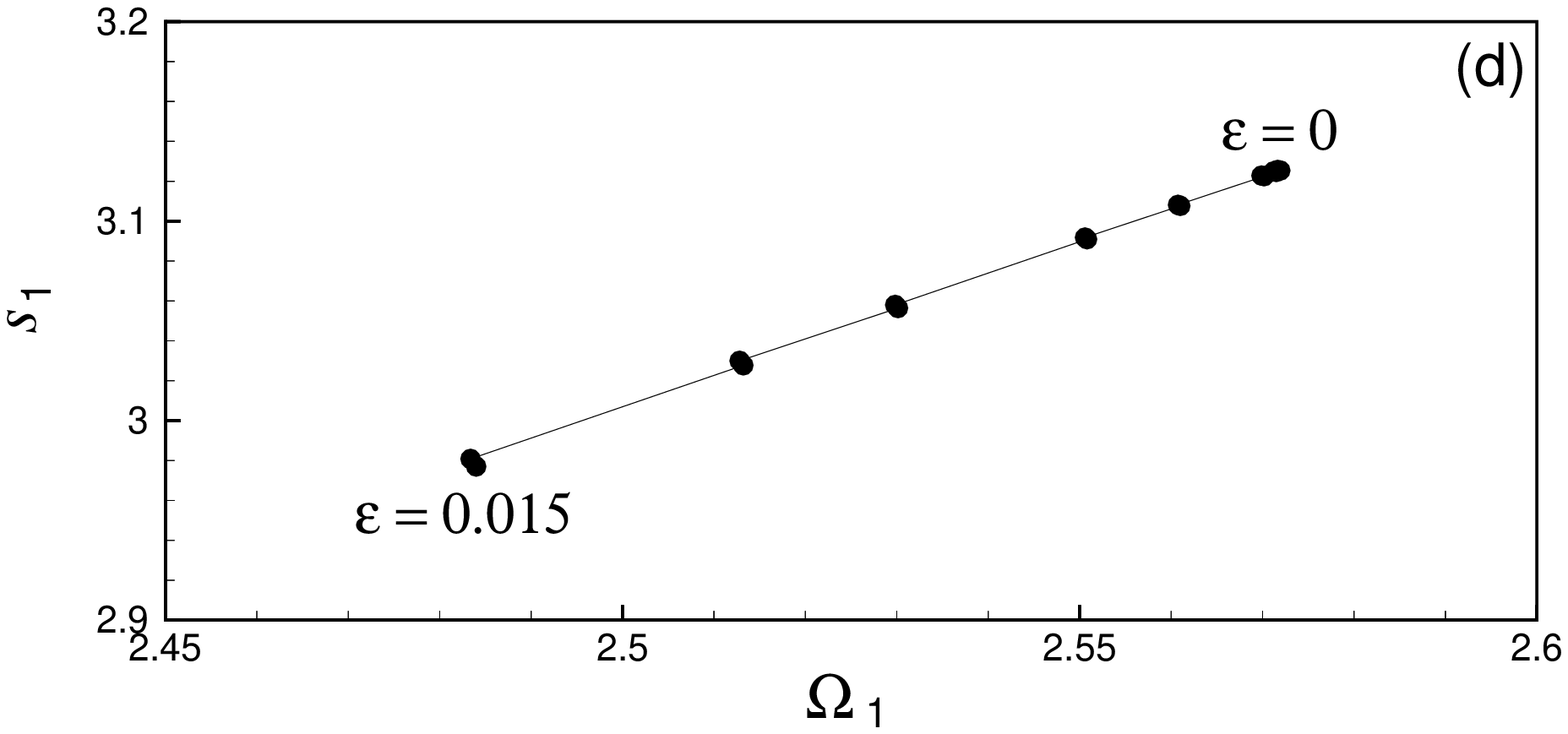}}}
\centerline{\hbox{\epsfxsize=8.0truecm\epsfbox{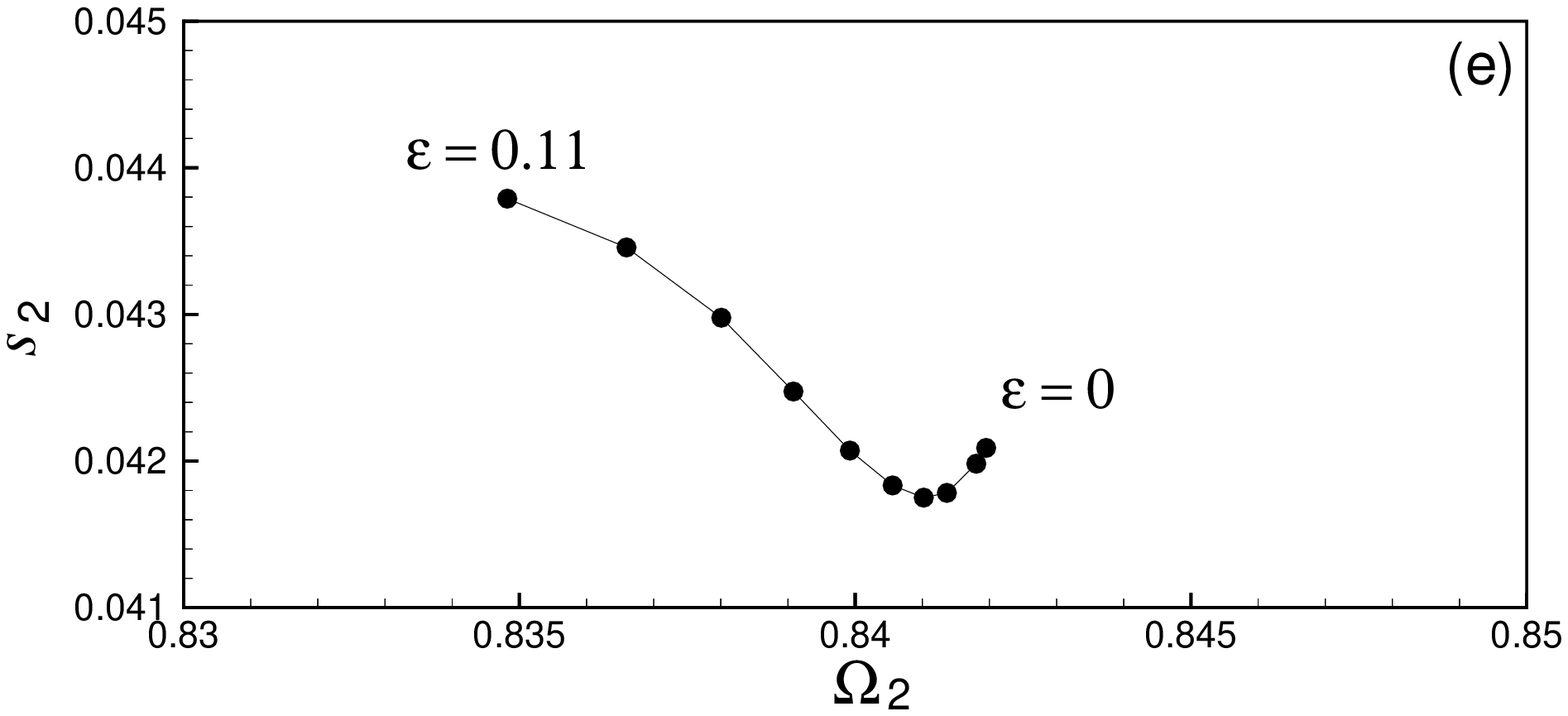}}
            \hspace{0.3truecm}
            \hbox{\epsfxsize=8.0truecm\epsfbox{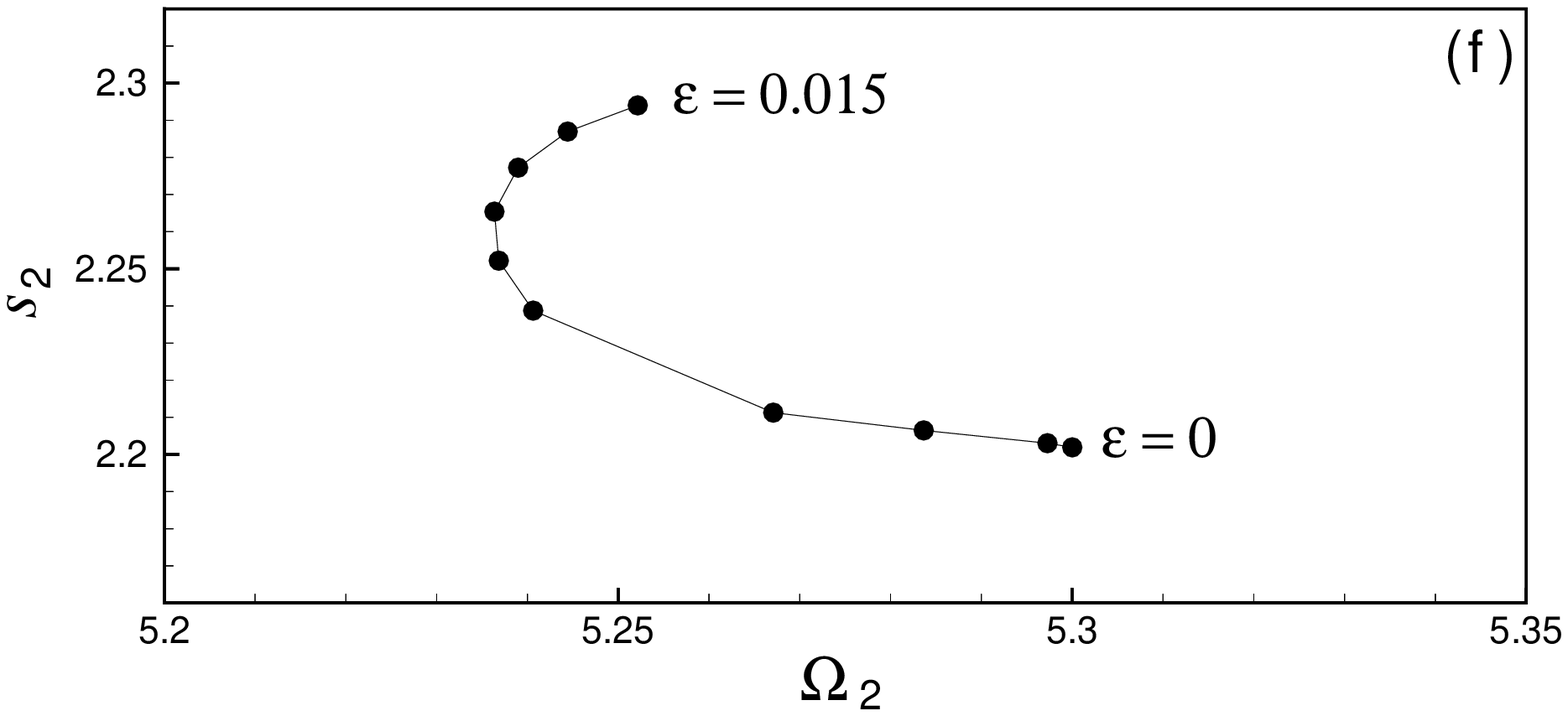}}}
\caption{Evolution of the eigenfrequencies of non-axisymmetric discs
with respect to $\epsilon$ and $\Mach$. Top, middle and bottom
panels correspond to $m=0$, 1 and 2, respectively. Right panels
correspond to $\Mach=5$. (a) $\Mach$=2.8 (c) $\Mach$=1.7 (e)
$\Mach=1.4$. As Panels {\em a} and {\em c} show, the $m=0$ and $m=1$
modes are stabilized by increasing $\epsilon$ in the allowable zone
of Fig. \ref{fig:fig1}. The $m=0$ mode can be stabilized for $2.7
\lta \Mach \lta 2.91$. This interval becomes $1.6 \lta \Mach \lta
2.3$ for $m=1$. The eigenfrequencies do not vary substantially for
cold discs with $\Mach=5$ and for the $m=2$ modes.} \label{fig:fig5}
\end{figure*}

\begin{figure*}
\centerline{\hbox{\epsfxsize=8.0truecm\epsfbox{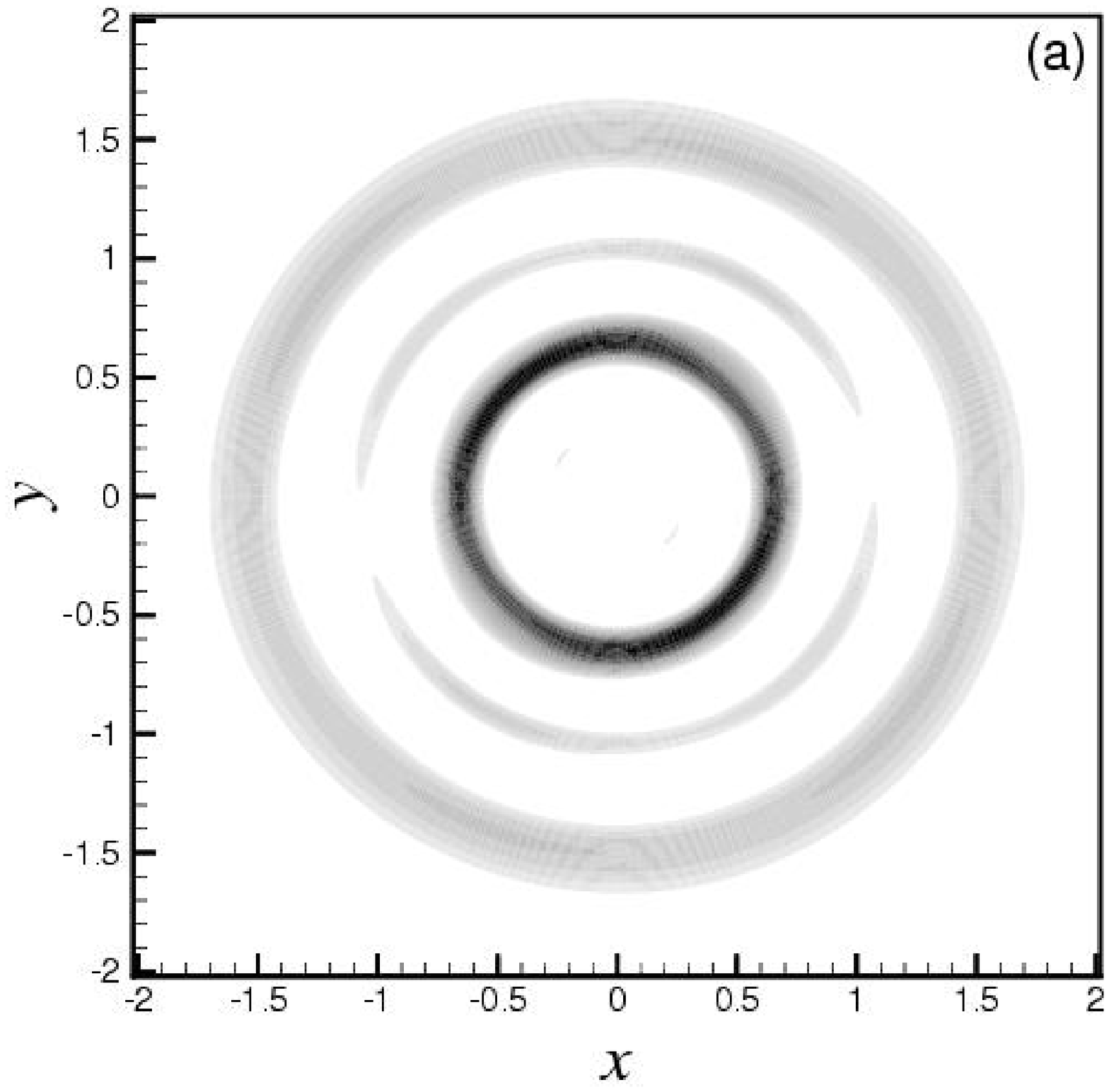}}
            \hspace{0.1truecm}
            \hbox{\epsfxsize=8.0truecm\epsfbox{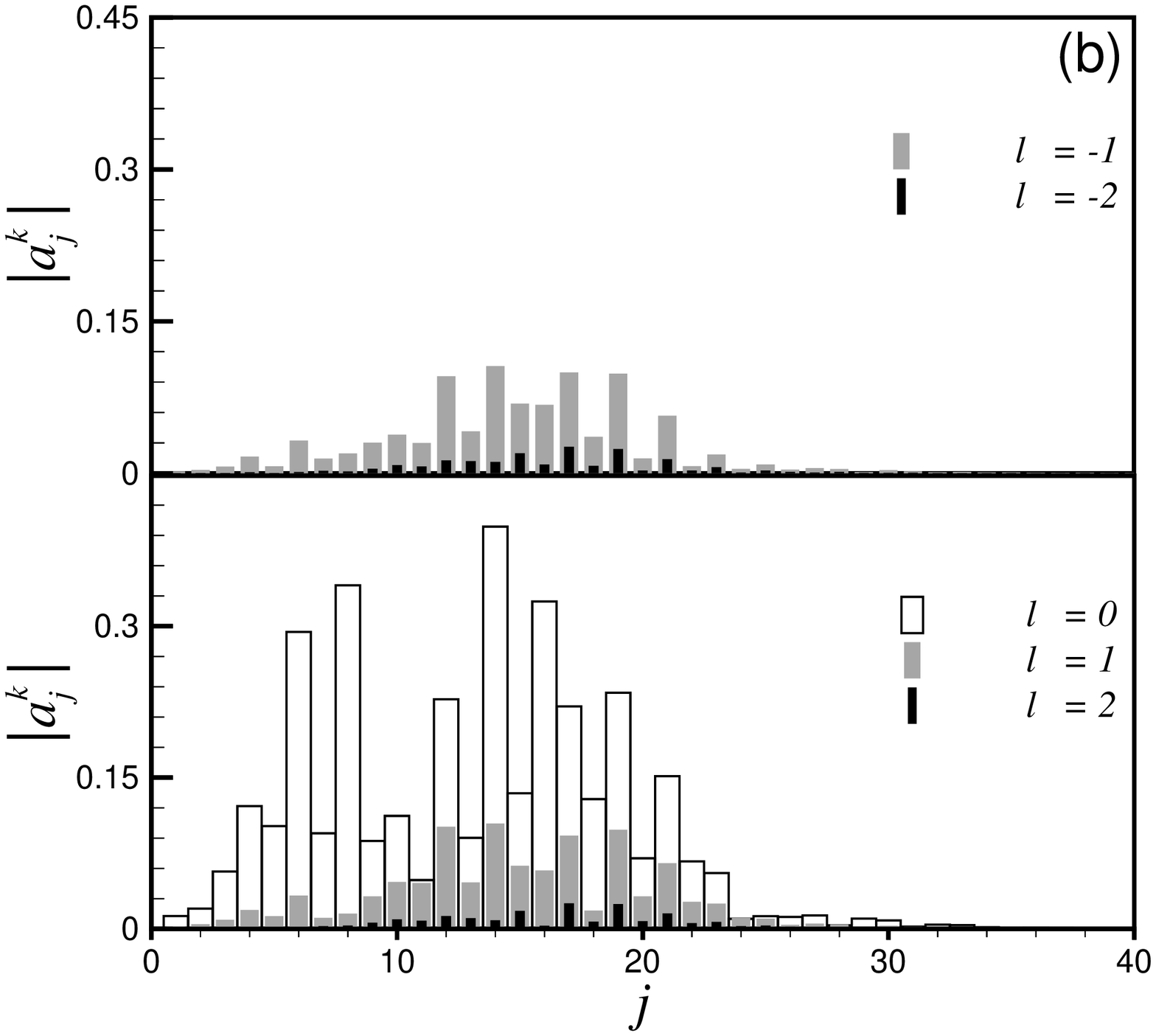}}}
\centerline{\hbox{\epsfxsize=8.0truecm\epsfbox{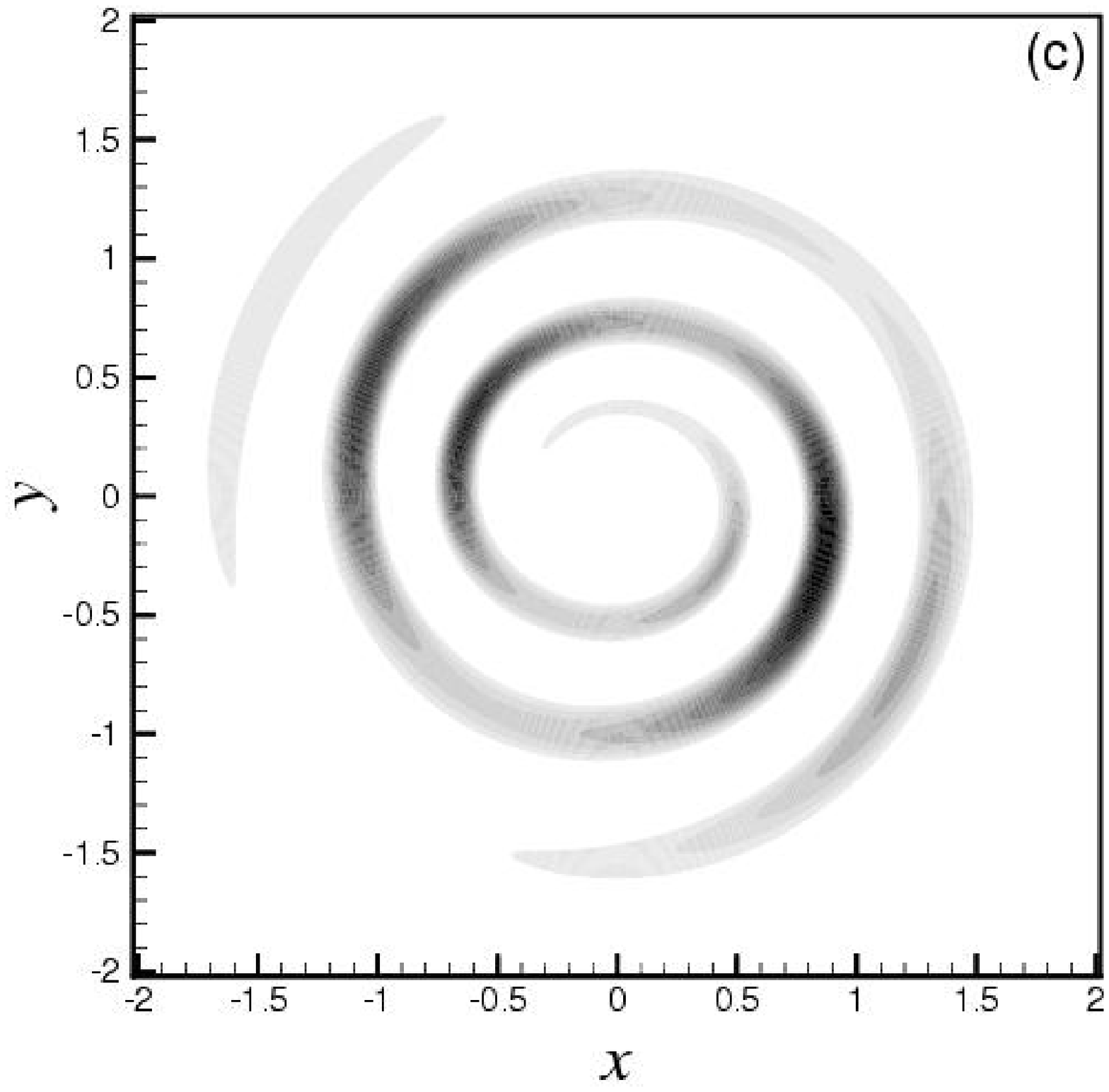}}
            \hspace{0.1truecm}
            \hbox{\epsfxsize=8.0truecm\epsfbox{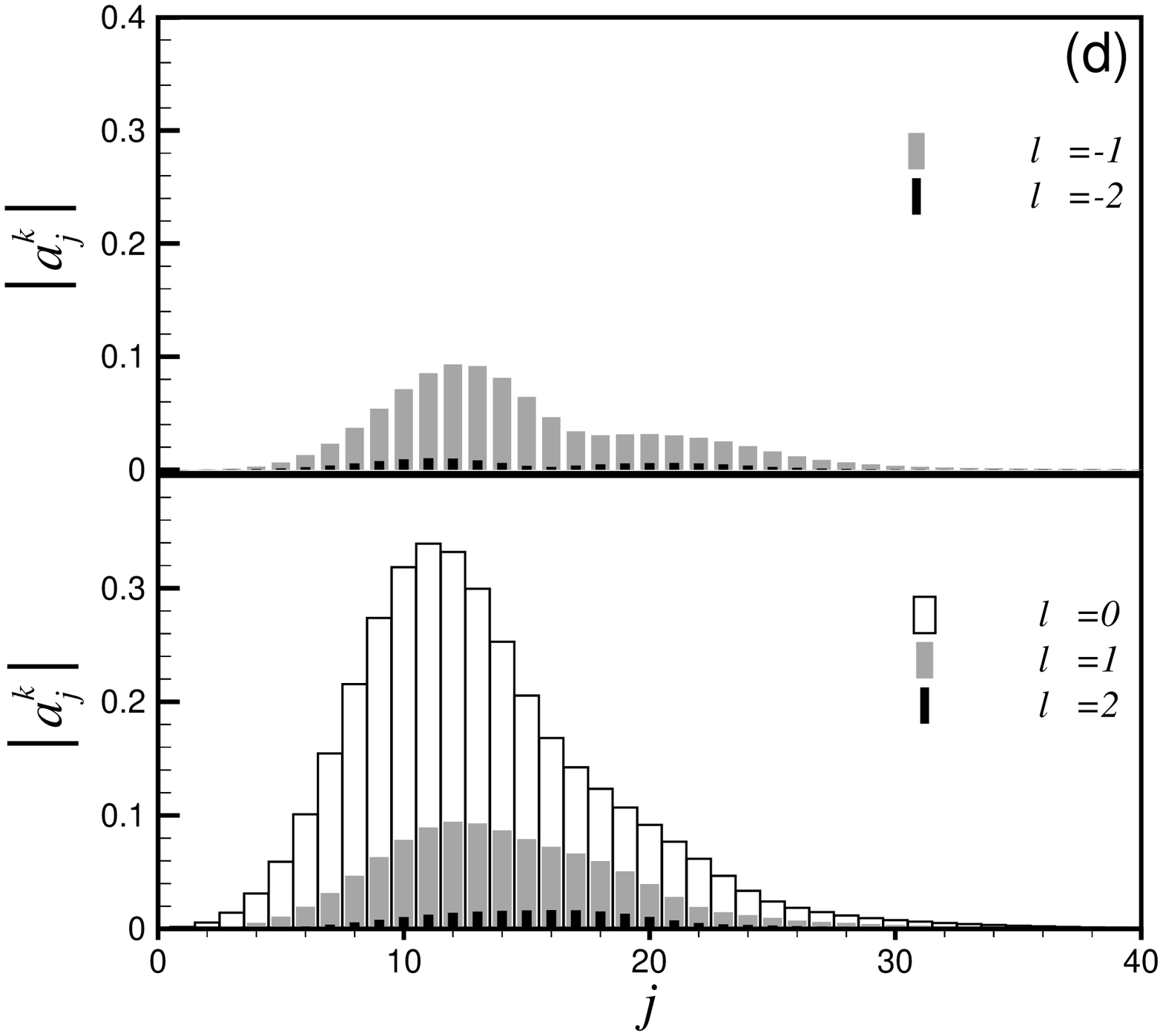}}}

\centerline{\hbox{\epsfxsize=8.0truecm\epsfbox{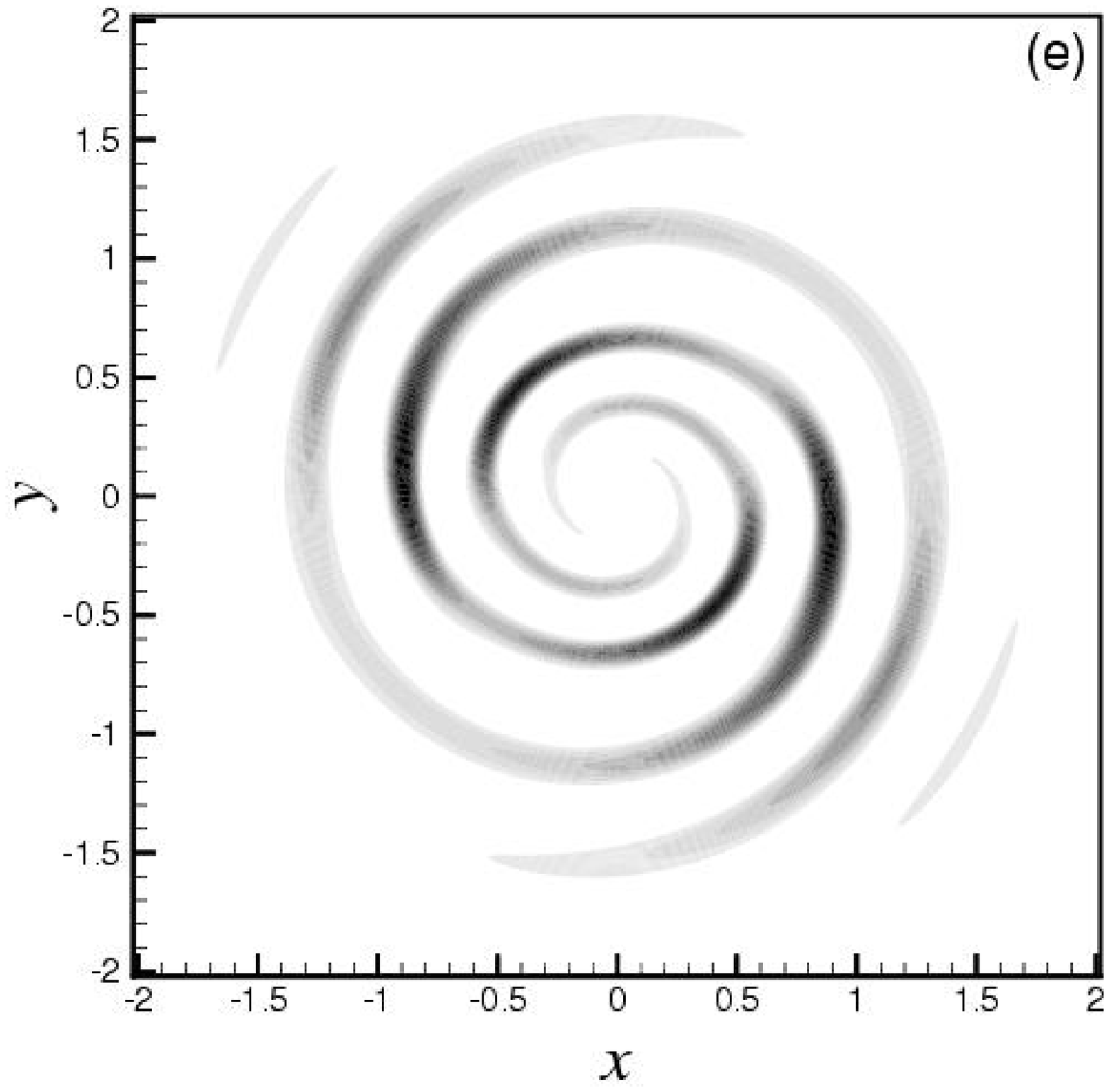}}
            \hspace{0.1truecm}
            \hbox{\epsfxsize=8.0truecm\epsfbox{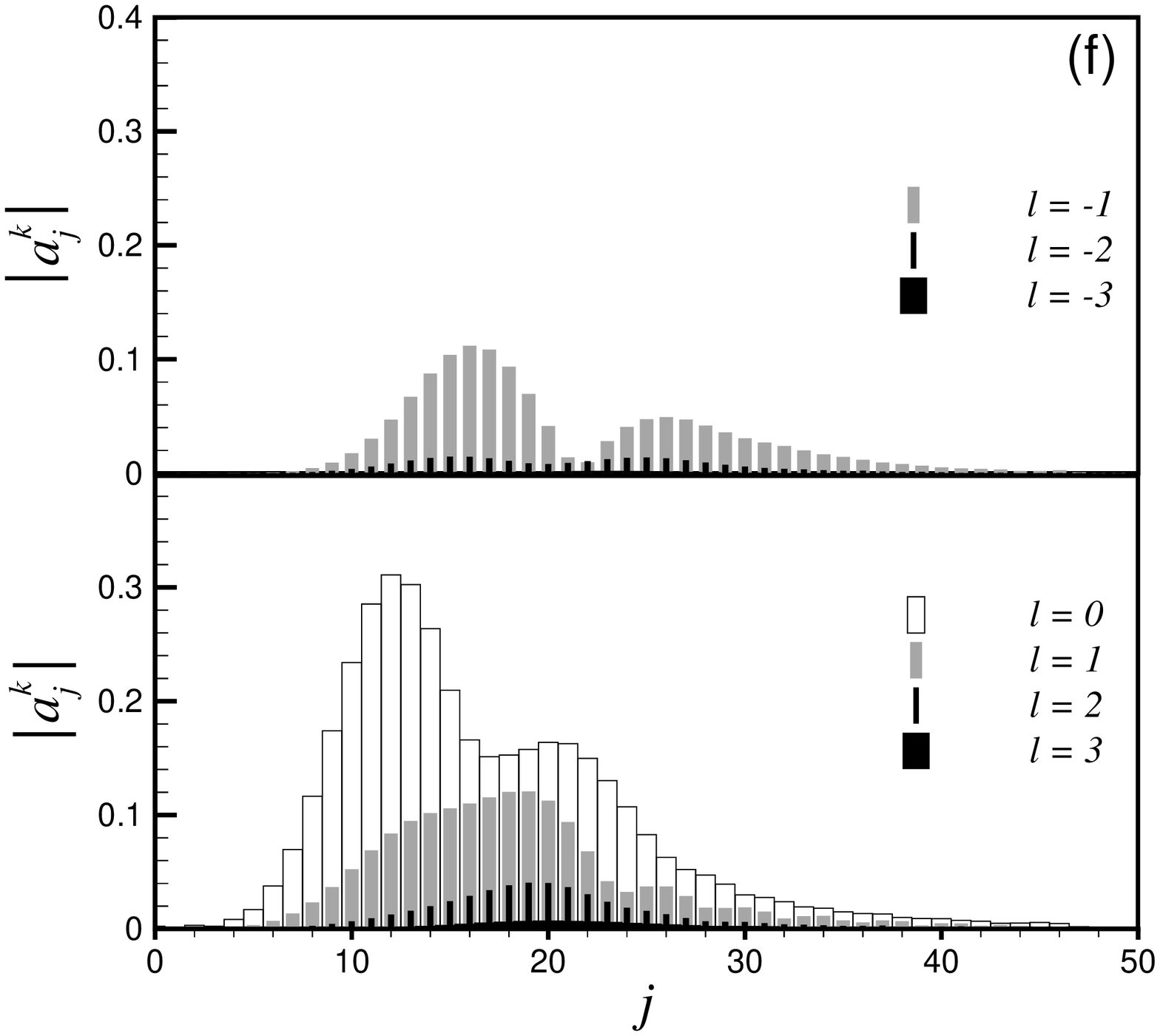}}}

%\centerline{\includegraphics[width=80mm]{fig6e.eps}
%            \hspace{0.1truecm}
%            \includegraphics[width=80mm]{fig6f.eps}}
%
\caption{Mode shapes (left panels) of a set of non-axisymmetric
discs with $\Mach=5.0$ and $\alpha=0.5$. Top, middle and bottom rows
correspond to $(m,\epsilon)=(0,0.015)$, (1,0.015) and (2,0.01),
respectively. These modes bifurcate from the modes of Fig.
\ref{fig:fig4}. Gray scale contours mark 10 to 100 percent of the
maximum perturbed density with a step of 10. Local density peaks
(darker regions) along the major spiral arms correspond to the local
maxima of $\Sigma^{(m)}_1(R,\phi,\epsilon)$. Right panel of each
mode shape shows the corresponding absolute value of
$a^k_j=\epsilon^{\vert l\vert}\hat a^k_j$ ($k=m+nl$) versus $j$ and
$l$. The results clearly show the convergence of our perturbation
series both in the radial and in the azimuthal directions.}
\label{fig:fig6}
\end{figure*}

Fig. \ref{fig:fig5} shows the evolution of the fundamental
eigenfrequencies with respect to the variations in $\epsilon$ for
two choices of $\Mach$ for each mode number $m$. Left panels in Fig.
\ref{fig:fig5} correspond to a marginal $\Mach$, which is chosen
slightly larger than the upper limit of Mach number for stable
circular discs. Right panels in Fig. \ref{fig:fig5} are for a cold
disc with $M=5$. Left panels in Fig. \ref{fig:fig6} display the
$m=0$, 1 and 2 modes of non-axisymmetric discs that bifurcate from
the modes of Fig. \ref{fig:fig4}. Gray scale contours show the
positive part of
\begin{equation}
\Sigma ^{(m)}_1(R,\phi,\epsilon)=A_m(R,\phi,\epsilon)\cos
[m\phi+\Theta_m(R,\phi,\epsilon)].
\end{equation}
Right panel to each mode shape shows the absolute values of the
scaled expansion coefficients, $\vert a^k_j\vert$=$\epsilon ^{\vert
l\vert}\vert \hat a^k_j\vert$ ($k=m+nl$), in terms of $l$ and $j$.


Our computations show that non-axisymmetric discs with $\Mach <2.7$
are stable for $m$=0 excitations independent of the magnitude of
$\epsilon$. In discs with $\Mach >2.7$ a tangent bifurcation occurs
in the complex $(\Omega_m,s_m)$-plane at the location of each
eigenfrequency (including $\omega_c$) and two sets of
eigenfrequencies with non-zero pattern speeds come to existence.
They are of the form $\omega_0$=$\pm \Omega_0 + {\rm i}s_0$. This is
an interesting result: we have found an unstable, ring-shaped
density perturbation that can rotate. Our computations show that the
models with $2.7 \lta \Mach \lta 2.91$ are stabilized by increasing
$\epsilon$. Fig. \ref{fig:fig5}{\em a} shows that how a model with
$\Mach$=2.8 is eventually stabilized for $\epsilon \approx 0.042$.
The pattern speed of the $m=0$ mode takes both positive and negative
values because equation (\ref{eq::perturbed-vec-m-mode}) is
invariant under the reflections $t\rightarrow -t$ and $\phi
\rightarrow -\phi$ for $m=0$. According to Fig. \ref{fig:fig1}, the
upper limit of $\epsilon$ decreases for our cold bisymmetric discs,
and therefore, models with large $\Mach$ cannot be stabilized only
by varying the non-axisymmetry parameter. Nonetheless, the placement
of eigenfrequencies in the complex plane is similar for all unstable
discs: the pattern rotates faster and grows slower as $\epsilon$
increases. Fig. \ref{fig:fig6}{\em a} displays the growing $m$=0
mode of a non-axisymmetric disc with $(\Mach,\epsilon)$=$(5,0.015)$.
The corresponding $\vert a^{m+nl}_j \vert$ (Fig. \ref{fig:fig6}{\em
b}) tend to zero as $\vert l \vert$ and $j$ are increased. Thus, the
perturbation expansion given in
(\ref{eq::perturbed-fields-expansion}) is convergent both in the
azimuthal and in the radial directions.

Our eigenfrequency calculations for $m=1$ has been summarized in
Figs \ref{fig:fig5}{\em c} and {\em d} for $M=$1.7 and $5$,
respectively. The unstable $m=1$ mode has a definite sense of
rotation and its pattern speed decreases as $\epsilon$ is increased.
For the disc with $\Mach=1.7$, $\omega_m$ steeply falls off and
ultimately vanishes for $\epsilon \approx 0.05$. This property is
shared by all models in the range $1.6 \lta \Mach \lta 2.3$. Models
with $\Mach < 1.6$ are stable for $m=1$ excitations regardless of
the magnitude of $\epsilon$. The fundamental eigenfrequency of the
models with $\Mach > 2.3$ does not vary considerably as $\epsilon$
is increased in the admissible zone of Fig. \ref{fig:fig1}. However,
the mode shape is highly influenced by the non-axisymmetry. Fig.
\ref{fig:fig6}{\em c} shows the unstable $m=1$ mode of a
non-axisymmetric disc with $(\Mach,\epsilon)$=$(5,0.015)$. Although
the overall shape is a single armed spiral, which corresponds to the
$l=0$ wave component of equation (\ref{eq::perturbations-for-all}),
a number of dense clumps occur along the major spiral arm. These
local clumps are generated by the $\vert l\vert
>0$ wave components of the perturbed density expansion. Fig.
\ref{fig:fig6}{\em d} shows the variation of $\vert a^{m+nl}_j\vert$
versus $l$ and $j$ for $m=1$. The sequence nicely converges as
$\vert l\vert$ is increased. It is seen that the choice of $L=2$
provides enough accuracy both for $m=0$ and for $m=1$.

Finding the eigenfrequencies of the $m=2$ mode for $\Mach <2$ is
troublesome because the convergent complex branch emerges very close
to divergent branches (see Fig. \ref{fig:fig2}{\em h}). We have
written a program that computes the full eigenvalue spectrum for
several choices of $J$ and searches between all eigenvalues for the
most convergent one. According to our program, convergent unstable
$m=2$ modes exist for $\Mach \ge 1.4$. Numerical results displayed
in Fig. \ref{fig:fig5}{\em e} and {\em f} show that the fundamental
eigenfrequency is robust to variations in $\epsilon$. The
corresponding mode shapes are double armed spirals with some sort of
dense clumps distributed along the major arms. The pitch angle of
the global arms deceases as the disc is cooled (large $\Mach$). Fig.
\ref{fig:fig6}{\em e} displays the $m=2$ mode of a non-axisymmetric
disc with $(\Mach,\epsilon)$=$(5,0.01)$. The perturbation series
(\ref{eq::perturbed-fields-expansion}) slowly converges in the
$\phi$-direction for $m=2$. So we kept more azimuthal wave
components and used $L=3$ (Fig. \ref{fig:fig6}{\em f}).

\section{Discussions}
\label{sec:discussion} The classical procedure in the instability
analysis of gaseous (and stellar) discs is to start with an
axisymmetric equilibrium and then perturb the governing equations in
the vicinity of this equilibrium state. The subsequent eigenvalue
problem [e.g., equation (\ref{eq:axisymmetric-eigenvalue-equation})]
results in a complex frequency $\omega_m =\Omega_m+{\rm i}s_m$,
which implies an $m$-fold barred/spiral pattern
$\Sigma_m=A_m(R)\cos[m\phi+\Theta_m(R)]$ that rotates by the angular
velocity $\Omega_m/m$ and exponentially grows by the rate $s_m$.
Nevertheless, most events of physical significance happen in
nonlinear regime when the modes begin to interact. In such a
circumstance, more than one azimuthal wave component is present in
the response spectrum. The most general form of a density wave in
nonlinear regime is
\begin{equation}
\Sigma_1(R,\phi,t)=\sum_{m=-\infty}^{+\infty}A_m(R,t)e^{{\rm
i}[m\phi+\Theta_m(R,t)]}, \label{eq::general-density-wave}
\end{equation}
which means that the amplitude and phase angle of $\Sigma_1$ are
functions of all spatial and time variables. A primary result of
(\ref{eq::general-density-wave}) is the generation of dense
clumps at the locations of the maxima of $\Sigma_1$. Clumps may
undergo gravitational collapse, merge and give birth to gas giants.
This is a widely accepted formation scenario of protoplanets (e.g.,
Boss 1997; Mayer et al. 2004) because it can take place in short
time scales. Asymmetries like the presence of a double star (Boss
2006) can also generate dense clumps.

According to (\ref{eq::mode-nonaxi}) and calculations of
\S\ref{sec:modes-nonaxi-discs}, unstable density waves of
non-axisymmetric discs have a rich structure (in linear regime) even
for very small non-axisymmetry parameter $\epsilon$. Although the
temporal evolution of the $m$th mode is governed by the simple
function $\exp({\rm i}\omega_m t)$, the eigenfunction associated
with $\omega_m$ is by no means simple and it contains infinite
number of azimuthal wave components, which in turn are capable of
generating local clumps. Since realistic accretion discs have not
initially the {\it nice} axisymmetry feature, their instability can
give birth to protoplanets sooner than what nonlinear hydrodynamical
simulations predict. A protoplanet, as a possible product of the
gravitational collapse of a dense clump, is separated from the rest
of the evolving gas and migrates until it is captured by a resonant
island in the phase space.

Our perturbation theory can be applied to any non-axisymmetric
configuration regardless of the nature of the centre. Cutouts, which
are special treatments designed for scale-free models, are not
needed in cored systems with finite total mass. The most important
factor in success of our perturbation theory is the smallness of the
non-axisymmetry parameter $\epsilon$, which prohibits the
interaction of azimuthal wave components in the zeroth order terms.
Our equilibrium density has a single azimuthal Fourier component
[see equation (\ref{eq:simple-disc})]. Accordingly, the harmonic
numbers of wave components differ in steps of $nl$
($l=0,\pm1,\pm2,\cdots$). For bisymmetric discs studied in this
paper, this means that wave components have only even or odd
harmonic numbers. We will have the most complete spectrum if the
equilibrium fields include Fourier terms with both even and odd
harmonic numbers, or if the disc is lopsided with $n=1$.

We presented three length scales to the perturbed equation
(\ref{eq::perturbations-for-all}). The first two are the cutout
radii, $R_{\rm in}$ and $R_{\rm out}$, and the third one is the
length scale of the basis functions, $R_C$. At a first look, such a
treatment of scale-free systems seems unfavorable because in this
way we destroy their scale-invariance and cuspy features. However,
with appropriate selection of the exponents of the cutout functions,
$N_{\rm in}$=$N_{\rm out}$=2, and the weight function of the
velocity components, $w_2(\xi)$, the coefficient matrix $\textbf{A}$
of our eigenvalue problem became a function of the two dimensionless
parameters $\alpha$ and $\gamma$. We easily decreased $\gamma$ and
noticed that unstable waves do not extend to distant regions
although the mass of our equilibrium models (with flat rotation
curves) is infinite at $R\rightarrow \infty$. We then let $\alpha$
tend to zero with the aim of removing the inner cutout. This needed
more terms of the series of associated Legendre functions, but the
qualitative features of our results remained unchanged as long as
mode shapes and ${\rm arg}(\omega_m)$ were concerned. There was an
excellent match between our results and those of Lemos et al. (1991)
who used logarithmic spirals to model their perturbed quantities.
Logarithmic spirals do not introduce any length scale to variational
equations.

The critical value of $\Mach$, below which our bisymmetric discs are
stable, is a function of the fundamental wave number $m$. Numerical
results show that the models with $\alpha =0.5$ are stable for $m=0$
excitations when $\Mach \lta 2.7$. This limit approximately becomes
$1.6$ and $1.4$ for $m=1$ and 2, respectively. The most interesting
property that we have found for unstable non-axisymmetric discs is
that the $m=0$ and $m=1$ modes can be stabilized by increasing the
non-axisymmetry parameter $\epsilon$. This happens for $m=0$ when
$2.7 \lta \Mach \lta 2.91$ and for $m=1$ when $1.6 \lta \Mach \lta
2.3$. There are some remarkable differences between the stabilized
$m=0$ and $m=1$ modes. The stabilized $m=0$ mode has the form of a
ring, which rotates according to the results displayed in Fig.
\ref{fig:fig5}{\em a}. It occurs when the wave components with the
azimuthal harmonic numbers $m-nl=-nl$ and $m+nl=nl$ cancel each
other for $l>0$. The stabilized $m=1$ mode, however, is a
non-spiral, equiangular neutral mode. It emerges for a sufficiently
large $\epsilon$, when the wave component with the harmonic number
$m-n=-1$ ($l=-1$) cancels the fundamental component with $m+n\times
0=1$ ($l=0$). It is worth noting that the $m=2$ mode cannot be
stabilized by increasing $\epsilon$, because the interacting wave
components of ${\cal O}(\epsilon)$ have the harmonic numbers $m-n=0$
($l=-1$) and $m+n=4$ ($l=1$). Neither of these first-order
components is capable of suppressing the effect of the fundamental
component with the harmonic number $m+n\times 0=2$. On the other
hand, the wave component with the harmonic number $m-n\times 2=-2$
is of ${\cal O}(\epsilon ^2)$, which is too weak to compensate the
zeroth order $m=2$ component. Based on the above reasonings, we
speculate that the $m=2$ mode can be stabilized in discs whose
surface density includes the fourth harmonic of the azimuthal angle
or when $n=4$ in the simple discs of JA. Furthermore, the unstable
$m=1$ and $m=2$ modes of lopsided discs (with $n=1$) cannot be
stabilized by increasing the non-axisymmetry parameter.

\section{Acknowledgments}
During the first 2 years of his PhD program, NMA received a research
grant from the Aerospace Research Institute through a collaboration
scheme between the IASBS and ARI. NMA also thanks Prof. Mohsen
Bahrami for insightful discussions.

\appendix

\section{\label{app::linear-operator}
         The elements of the Linear operator ${\bf L}$}

The linear operator ${\bf L}$ appeared in equation
(\ref{eq:linearized-perturbed-equations}) is in terms of the
active surface density $\Sigma _{\rm act}(R,\phi)$, and
the velocity field
\begin{equation}
\left (u_{20},u_{30} \right )$=$\left [v_{R0}(R,\phi),v_{\phi
0}(R,\phi)\right ],
\end{equation}
of the equilibrium state. The elements of ${\bf L}$ are
\begin{eqnarray}
L_{11} \!\!\!\! &=& \!\!\!\! {1 \over \Sigma_{\rm act}} \bigg (
{\partial _t} + {\vr \over R}+\vr \partial _R+ {\vp \over
R}\partial _\phi \nonumber \\ \!\!\!\! &{}& \!\!\!\! \qquad\qquad
+\partial _R \vr + {\partial _\phi \vp \over R} \bigg ),
\\
L_{12} \!\!\!\! &=& \!\!\!\! {1\over R}+ \left [ \partial _R +
\partial _R \left ( \ln \Sigma _{\rm act} \right ) \right ], \\
L_{13} \!\!\!\! &=& \!\!\!\! {1\over R} \left [ \partial _\phi +
\partial _\phi \left ( \ln \Sigma _{\rm act} \right ) \right ], \\
L_{21} \!\!\!\! &=& \!\!\!\! {c^2(\phi)\over \Sigma_{\rm act}}
\left [ \partial _R - \partial _R \left ( \ln \Sigma _{\rm act}
\right ) \right ], \\
L_{22} \!\!\!\! &=& \!\!\!\! {\partial _t}+\partial _R v_{R0}
+v_{R0}
\partial _R + {v_{\phi 0}\over R}\partial _\phi, \\
L_{23} \!\!\!\! &=& \!\!\!\! {1\over R}\partial _\phi v_{R0} -
{2 v_{\phi 0} \over R}, \\
L_{31} \!\!\!\! &=& \!\!\!\! {1\over \Sigma_{\rm act}}
\left \{ \frac 1R \partial _\phi c^2(\phi) + {c^2(\phi)\over R}
\left [ \partial _\phi - \partial _\phi \left ( \ln \Sigma_{\rm act}
\right ) \right ] \right \}, \\
L_{32} \!\!\!\! &=& \!\!\!\! \partial _R v_{\phi 0}+ {v_{\phi
0}\over R}, \\ L_{33} \!\!\!\! &=& \!\!\!\! {\partial _t} +\frac
1{R}\partial _\phi v_{\phi 0}+{v_{R 0}\over R}+v_{R 0}\partial _R
+{v_{\phi 0}\over R}
\partial _\phi.
\end{eqnarray}

\section{The elements of the operator matrices $\textbf{D}_p$}
\label{app::dp-elements}

The operators ${\textbf{D}}_{p}$=$[D_{p,ij}]$ ($i,j=1,2,3$) appear
in the interacting terms of (\ref{eq::perturbed-vec-m-mode}). They
are functions of $R$ and ${\rm d}/{\rm d}R$. Let us define $k=m+nl$
and introduce
\begin{eqnarray}
\!\!\!&{}&\!\!\! H(R) = \frac{R^2}{R^2+\alpha^2 R_0^2}\
\frac{R_0^2}{\gamma^2 R^2+R_0^2},
\nonumber\\
\!\!\!&{}&\!\!\! H_{1}(R) =
1+\alpha^2\gamma^2+3\gamma^2\left(\frac{R}{R_0}\right)^2-\alpha^2
\left(\frac{R_0}{R}\right)^2,
\nonumber\\
\!\!\!&{}&\!\!\! \mu_1
=1-\left[\frac{5\Mach^2-1+3n(\Mach^2-1)}{2(1+n)}\right]\epsilon^2,
\nonumber\\
\!\!\!&{}&\!\!\! \mu_2 = \frac{4+3\epsilon^2+8n(1-2n)\left
(1+\epsilon^2 \right )}{8(1+n)(-1+4n^2)} \nonumber \\ \!\!\! &{}&
\!\!\! -\frac{16n^3\left (2+3\epsilon^2 \right )}{8(1+n)(-1+4n^2)}
\nonumber\\
\!\!\!&{}&\!\!\!+\Mach^2\left[\frac{2+n}{2(1+n)}+
\frac{72n^3+104n^2-22n-27}{8(1+n)(-1+4n^2)} \ \epsilon^2\right],
\nonumber\\
\!\!\!&{}&\!\!\! \mu_3 = \frac{12n^3+6n^2-n-1}{4(1+n)(-1+4n^2)}
\nonumber\\
\!\!\!&{}&\!\!\!-\Mach^2\frac{12n^3+18n^2-5n-5}{4(1+n)(-1+4n^2)},
\nonumber\\
\!\!\!&{}&\!\!\! \mu_4 = \frac{-16n^3-8n^2+1}{8(1+n)(-1+4n^2)}
\nonumber\\
\!\!\!&{}&\!\!\!+\Mach^2\frac{24n^3+32n^2-10n-9}{8(1+n)(-1+4n^2)}.
\nonumber
\end{eqnarray}
For $\vert p\vert \le 3$, we have found
\begin{eqnarray}
D_{0,11}\!\!\!&=&\!\!\! \textrm{i} k \Mach
\left(\frac{R_0}{R}\right)\left (1+\frac12 \epsilon^2 \right ),
\nonumber\\
D_{0,12}\!\!\!&=&\!\!\! 2
H^2(R)\left[\alpha^2\left(\frac{R_0}{R}\right)^4-\gamma^2\right]\nonumber
\\
\!\!\!&{}&\!\!\!+H(R)\left(\frac{R_0^2}{R}\right)\frac{d}{dR},
\nonumber\\
D_{0,13}\!\!\!&=&\!\!\! \textrm{i} k
H(R)\left(\frac{R_0}{R}\right)^2,
\nonumber\\
D_{0,21}\!\!\!&=&\!\!\! \mu_1 H_{1}(R)+\mu_1 \frac{R}{H(R)}
\frac{d}{dR},
\nonumber\\
D_{0,22}\!\!\!&=&\!\!\! \textrm{i} k
\Mach\left(\frac{R_0}{R}\right)\left (1+\frac12 \epsilon^2 \right ),
\nonumber \\
D_{0,23}\!\!\!&=&\!\!\! -2 \Mach\left(\frac{R_0}{R}\right)\left
(1+\frac12 \epsilon^2 \right ),
\nonumber \\
D_{0,31}\!\!\!&=&\!\!\! \textrm{i} k \mu_1
\frac{1}{H(R)},\nonumber \\
D_{0,32}\!\!\!&=&\!\!\! \Mach\left(\frac{R_0}{R}\right)\left
(1+\frac12 \epsilon^2 \right ),
\nonumber\\
D_{0,33}\!\!\!&=&\!\!\! \textrm{i} k \Mach
\left(\frac{R_0}{R}\right)\left (1+\frac12 \epsilon^2 \right ),
\nonumber\\
D_{+1,11}\!\!\!&=&\!\!\! -\textrm{i} \frac{1+\Mach^2}{2\Mach(1+n)}
\nonumber
\\ \!\!\! &{}&\!\!\!
\times
\left[H_1(R)H(R)\left(\frac{R_0}{R}\right)+R_0\frac{d}{dR}\right]
\nonumber\\
\!\!\!&{}&\!\!\!-\textrm{i} k\Mach \left (\frac{1}{2}+\frac38
\epsilon^2 \right )\left(\frac{R_0}{R}\right),
\nonumber\\
D_{+1,12}\!\!\!&=&\!\!\!
H^2(R)\left[\alpha^2\left(\frac{R_0}{R}\right)^4-\gamma^2\right]\nonumber\\
\!\!\!&{}&\!\!\!+\frac12H(R)\left(\frac{R_0^2}{R}\right)\frac{d}{dR},
\nonumber\\
D_{+1,13}\!\!\!&=&\!\!\! \frac12\textrm{i}k
H(R)\left(\frac{R_0}{R}\right)^2,
\nonumber\\
D_{+1,21}\!\!\!&=&\!\!\! \mu_2 H_1(R) +\mu_2 \frac{R}{H(R)}
\frac{d}{dR},
\nonumber\\
D_{+1,22}\!\!\!&=&\!\!\! -\textrm{i}
\frac{1+\Mach^2}{2\Mach(1+n)}R_0\frac{d}{dR}\nonumber\\
\!\!\!&{}&\!\!\!-\textrm{i} \Mach(k+n)\left (\frac{1}{2}+\frac38
\epsilon^2 \right )\left(\frac{R_0}{R}\right),
\nonumber\\
D_{+1,23}\!\!\!&=&\!\!\!
\left[\frac{\Mach^2(2+n)-n}{2\Mach(1+n)}+\frac34\Mach
\epsilon^2\right] \left(\frac{R_0}{R}\right),
\nonumber\\
D_{+1,31}\!\!\!&=&\!\!\! \textrm{i}k \mu_2 \frac{1}{H(R)},
\nonumber\\
D_{+1,32}\!\!\!&=&\!\!\! -\Mach \left (\frac{1}{2}+\frac38
\epsilon^2 \right ) \left(\frac{R_0}{R}\right),
\nonumber\\
D_{+1,33}\!\!\!&=&\!\!\! -\textrm{i}
\frac{1+\Mach^2}{2\Mach(1+n)}\left[R_0\frac{d}{dR}+
\left(\frac{R_0}{R}\right)\right]\nonumber\\
\!\!\!&{}&\!\!\!-\textrm{i} k \Mach \left (\frac{1}{2}+\frac38
\epsilon^2 \right ) \left(\frac{R_0}{R}\right),
\nonumber\\
D_{-1,11}\!\!\!&=&\!\!\! \textrm{i} \frac{1+\Mach^2}{2\Mach(1+n)}
\nonumber
\\
\!\!\! &{}& \!\!\! \times
\left[H_1(R)H(R)\left(\frac{R_0}{R}\right)+R_0\frac{d}{dR}\right]
\nonumber\\
\!\!\!&{}&\!\!\!-\textrm{i} k\Mach \left (\frac{1}{2}+\frac38
\epsilon^2 \right )\left(\frac{R_0}{R}\right),
\nonumber\\
D_{-1,12}\!\!\!&=&\!\!\!
H^2(R)\left[\alpha^2\left(\frac{R_0}{R}\right)^4-\gamma^2\right]\nonumber\\
\!\!\!&{}&\!\!\!+\frac12H(R)\left(\frac{R_0^2}{R}\right)\frac{d}{dR},
\nonumber\\
D_{-1,13}\!\!\!&=&\!\!\! \frac12\textrm{i}k
H(R)\left(\frac{R_0}{R}\right)^2,
\nonumber\\
D_{-1,21}\!\!\!&=&\!\!\! \mu_2 H_1(R) +\mu_2 \frac{R}{H(R)}
\frac{d}{dR},
\nonumber\\
D_{-1,22}\!\!\!&=&\!\!\! \textrm{i}
\frac{1+\Mach^2}{2\Mach(1+n)}R_0\frac{d}{dR}\nonumber\\
\!\!\!&{}&\!\!\!-\textrm{i} \Mach(k-n) \left (\frac{1}{2}+\frac38
\epsilon^2 \right )\left(\frac{R_0}{R}\right),
\nonumber\\
D_{-1,23}\!\!\!&=&\!\!\!
\left[\frac{\Mach^2(2+n)-n}{2\Mach(1+n)}+\frac34\Mach
\epsilon^2\right] \left(\frac{R_0}{R}\right),
\nonumber \\
D_{-1,31}\!\!\!&=&\!\!\! \textrm{i}k \mu_2 \frac{1}{H(R)},
\nonumber \\
D_{-1,32}\!\!\!&=&\!\!\! -\Mach \left (\frac{1}{2}+\frac38
\epsilon^2 \right ) \left(\frac{R_0}{R}\right)
\nonumber \\
D_{-1,33}\!\!\!&=&\!\!\! \textrm{i}
\frac{1+\Mach^2}{2\Mach(1+n)}\left[R_0\frac{d}{dR}+\left(\frac{R_0}{R}\right)\right]\nonumber\\
\!\!\!&{}&\!\!\!-\textrm{i} k \Mach \left (\frac{1}{2}+\frac38
\epsilon^2 \right ) \left(\frac{R_0}{R}\right),
\nonumber \\
D_{+2,11}\!\!\!&=&\!\!\! \textrm{i}
\frac{1+\Mach^2}{4\Mach(-1+4n^2)} \nonumber \\ \!\!\! &{}& \!\!\!
\times
\left[H_1(R)H(R)\left(\frac{R_0}{R}\right)+R_0\frac{d}{dR}\right]
\nonumber \\
\!\!\!&{}&\!\!\! + \frac{1}{4}\textrm{i} k
\Mach\left(\frac{R_0}{R}\right),
\nonumber \\
D_{+2,12}\!\!\!&=&\!\!\! 0
\nonumber \\
D_{+2,13}\!\!\!&=&\!\!\! 0
\nonumber \\
D_{+2,21}\!\!\!&=&\!\!\! \mu_3 H_1(R) +\mu_3
\frac{R}{H(R)}\frac{d}{dR}
\nonumber \\
D_{+2,22}\!\!\!&=&\!\!\!
\textrm{i}\frac{1+\Mach^2}{4\Mach(-1+4n^2)}R_0\frac{d}{dR}\nonumber
\\ \!\!\! &{}& \!\!\! +\textrm{i}\frac{\Mach(k+2n)}{4}\left(\frac{R_0}{R}\right)
\nonumber \\
D_{+2,23}\!\!\!&=&\!\!\!
\frac{n+\Mach^2(1+n-4n^2)}{2\Mach(-1+4n^2)}\left(\frac{R_0}{R}\right)
\nonumber \\
D_{+2,31}\!\!\!&=&\!\!\! \textrm{i}k \mu_3 \frac{1}{H(R)}
\nonumber \\
D_{+2,32}\!\!\!&=&\!\!\! \frac{1}{4}\Mach\left(\frac{R_0}{R}\right)
\nonumber \\
D_{+2,33}\!\!\!&=&\!\!\!
\textrm{i}\frac{1+\Mach^2}{4\Mach(-1+4n^2)}
\left[\left(\frac{R_0}{R}\right)+R_0\frac{d}{dR}\right]\nonumber\\
\!\!\!&{}&\!\!\!+\frac{1}{4}\textrm{i}k \Mach
\left(\frac{R_0}{R}\right)
\nonumber \\
D_{-2,11}\!\!\!&=&\!\!\! -\textrm{i}
\frac{1+\Mach^2}{4\Mach(-1+4n^2)} \nonumber
\\ \!\!\! &{}& \!\!\! \times
\left[H_1(R)H(R)\left(\frac{R_0}{R}\right)+R_0\frac{d}{dR}\right]
\nonumber \\
\!\!\!&{}&\!\!\! + \frac{1}{4}\textrm{i} k
\Mach\left(\frac{R_0}{R}\right),
\nonumber \\
D_{-2,12}\!\!\!&=&\!\!\! 0,
\nonumber \\
D_{-2,13}\!\!\!&=&\!\!\! 0,
\nonumber \\
D_{-2,21}\!\!\!&=&\!\!\! \mu_3 H_1(R) +\mu_3
\frac{R}{H(R)}\frac{d}{dR},
\nonumber \\
D_{-2,22}\!\!\!&=&\!\!\!
-\textrm{i}\frac{1+\Mach^2}{4\Mach(-1+4n^2)}R_0\frac{d}{dR}\nonumber
\\ \!\!\! &{}& \!\!\! +\textrm{i}\frac{\Mach(k-2n)}{4}
\left(\frac{R_0}{R}\right),
\nonumber \\
D_{-2,23}\!\!\!&=&\!\!\!
\frac{n+\Mach^2(1+n-4n^2)}{2\Mach(-1+4n^2)}\left(\frac{R_0}{R}\right),
\nonumber\\
D_{-2,31}\!\!\!&=&\!\!\! \textrm{i}k \mu_3 \frac{1}{H(R)},
\nonumber\\
D_{-2,32}\!\!\!&=&\!\!\! \frac{1}{4}\Mach\left(\frac{R_0}{R}\right),
\nonumber\\
D_{-2,33}\!\!\!&=&\!\!\!-\textrm{i}\frac{1+\Mach^2}{4\Mach(-1+4n^2)}
\left[\left(\frac{R_0}{R}\right)+R_0\frac{d}{dR}\right]\nonumber\\
\!\!\!&{}&\!\!\!+\frac{1}{4}\textrm{i}k \Mach
\left(\frac{R_0}{R}\right),
\nonumber\\
D_{+3,11}\!\!\!&=&\!\!\! -\frac18\textrm{i} k \Mach
\left(\frac{R_0}{R}\right),
\nonumber\\
D_{+3,12}\!\!\!&=&\!\!\! 0,
\nonumber\\
D_{+3,13}\!\!\!&=&\!\!\! 0,
\nonumber\\
D_{+3,21}\!\!\!&=&\!\!\!\mu_4 H_1(R) +\mu_4
\frac{R}{H(R)}\frac{d}{dR},
\nonumber\\
D_{+3,22}\!\!\!&=&\!\!\!
-\frac18\textrm{i}\Mach(k+3n)\left(\frac{R_0}{R}\right),
\nonumber\\
D_{+3,23}\!\!\!&=&\!\!\! \frac{1}{4}\Mach\left(\frac{R_0}{R}\right),
\nonumber\\
D_{+3,31}\!\!\!&=&\!\!\! \textrm{i}k \mu_4 \frac{1}{H(R)},
\nonumber\\
D_{+3,32}\!\!\!&=&\!\!\!
-\frac{1}{8}\Mach\left(\frac{R_0}{R}\right),
\nonumber\\
D_{+3,33}\!\!\!&=&\!\!\!
-\frac{1}{8}\textrm{i}k\Mach\left(\frac{R_0}{R}\right),
\nonumber\\
D_{-3,11}\!\!\!&=&\!\!\!
-\frac{1}{8}\textrm{i}k\Mach\left(\frac{R_0}{R}\right),
\nonumber\\
D_{-3,12}\!\!\!&=&\!\!\! 0,
\nonumber\\
D_{-3,13}\!\!\!&=&\!\!\! 0,
\nonumber\\
D_{-3,21}\!\!\!&=&\!\!\! \mu_4 H_1(R) +\mu_4
\frac{R}{H(R)}\frac{d}{dR},
\nonumber\\
D_{-3,22}\!\!\!&=&\!\!\!
-\frac{1}{8}\textrm{i}\Mach(k-3n)
\left(\frac{R_0}{R}\right),
\nonumber\\
D_{-3,23}\!\!\!&=&\!\!\! \frac{1}{4}
\Mach\left(\frac{R_0}{R}\right),
\nonumber\\
D_{-3,31}\!\!\!&=&\!\!\! \textrm{i}k \mu_4 \frac{1}{H(R)},
\nonumber\\
D_{-3,32}\!\!\!&=&\!\!\!
-\frac{1}{8}\Mach\left(\frac{R_0}{R}\right),
\nonumber\\
D_{-3,33}\!\!\!&=&\!\!\!
-\frac{1}{8}\textrm{i}k\Mach\left(\frac{R_0}{R}\right).
\nonumber\\
\end{eqnarray}

\bsp \label{lastpage}

\end{document}